\documentclass[12pt]{article}
\usepackage{amssymb}
\usepackage{amsmath}
\usepackage{graphicx}
\usepackage{indentfirst}
\usepackage{cite}

\allowdisplaybreaks

\begin{document}

\title{Cross sections for 2-to-3 meson-meson scattering}
\author{Wan-Xia Li$^1$, Xiao-Ming Xu$^1$, and H. J. Weber$^2$}
\date{}
\maketitle \vspace{-1cm}
\centerline{$^1$Department of Physics, Shanghai University, Baoshan,
Shanghai 200444, China}
\centerline{$^2$Department of Physics, University of Virginia, Charlottesville,
VA 22904, USA}

\begin{abstract}
We study 2-to-3 meson-meson scattering based on the process that a gluon is 
created from a constituent quark or antiquark and subsequently the gluon 
creates 
a quark-antiquark pair. The transition potential for the process is derived in
QCD. Eight Feynman diagrams at tree level are involved in the 2-to-3
meson-meson scattering. Starting from the $S$-matrix element, we derive
the unpolarized cross section from the eight transition amplitudes 
corresponding
to the eight Feynman diagrams. The transition amplitudes contain 
color, spin, and flavor matrix elements. The 2-to-3 meson-meson scattering
includes 
$\pi \pi \to \pi K\bar{K}$,~$\pi K \to \pi \pi K$,~$\pi K \to KK\bar{K}$,
~$KK \to \pi KK$, and $K\bar{K} \to \pi K\bar{K}$.
Cross sections for the reactions are calculated. The cross sections depend on 
temperature obviously, and the
cross section for $\pi K \to \pi\pi K$ for total isospin $I=3/2$ at zero 
temperature is compared to experimental data.
By comparison with inelastic 2-to-2 meson-meson scattering, 
we find that 2-to-3 meson-meson scattering may be as important as 
inelastic 2-to-2 meson-meson scattering.
\end{abstract}



\vspace{0.5cm}
\leftline{\bf I. INTRODUCTION}
\vspace{0.5cm}

Many experiments and analyses have been done for 
elastic $\pi \pi$ scattering, elastic $\pi K$ scattering, and elastic 
$K\bar K$ scattering. Elastic phase shifts and elastic cross sections for 
$\pi \pi$ 
scattering have been measured via $\pi^- p \to \pi^- \pi^- \Delta^{++}$ 
\cite{Alitti,CMSJS,LCFMP}, $\pi^+ p \to \pi^0 \pi^0 \Delta^{++}$ \cite{CCBBB},
$\pi^+ p \to \pi^+ \pi^+ n$ \cite{ABBBH,Hoogland,AKMMP}, 
$\pi^- p \to \pi^0 \pi^0 n$ \cite{Takamatsu},
$\pi^- d \to \pi^- \pi^- pp$ \cite{CFSW,DBGGT}, 
$\pi^+ p \to \pi^+ \pi^- \Delta^{++}$ \cite{PABFF},
$\pi^{\pm} p \to \pi^{\pm} \pi^0 p$ \cite{BLR,Walker,AKMMP}, 
$\pi^- p \to \pi^- \pi^+ n$  
\cite{AKMMP,OGMWP,Hyams,Grayer,EM,Srinivasan,FP,BBMPS,KLR}, 
and $K^{\pm} \to \pi^+ \pi^- e^{\pm} \nu$ \cite{Rosselet,Pislak,GKPEY,Batley}. 
When the total isospin of the two pions is 0 or 1,
from phase shifts and cross sections one can identify resonances such 
as $f_0(980)$, $f_2(1270)$, $f_0(1370)$, $a_0(980)$, and $\rho (1450)$.
When the total isospin equals 2, the cross section for elastic $\pi \pi$
scattering can go up to 12 mb.
Elastic phase shifts and elastic cross sections for $\pi K$ 
scattering have been measured via $K^{\pm} p \to K^{\pm} \pi^- \Delta^{++}$ 
\cite{Mercer,JMTVH,Linglin,ECMDL}, $K^+ p \to K^0 \pi^0 \Delta^{++}$
\cite{Mercer}, $K^{\pm} p \to K^{\pm} \pi^+ n$ \cite{ECMDL,Aston}, and
$D^+ \to K^- \pi^+ e^+ \nu_e$ \cite{Sanchez}.
From elastic $\pi K$ scattering with a total isospin of 1/2,
one can find resonances such as $K_0^*(1430)$.
The cross section for elastic $\pi K$ scattering with the other total isospin 
3/2 may become as large as 4.6 mb. The elastic cross section
for $K^+ K^-$ scattering obtained from $\pi^+ p \to K^+ K^- \Delta^{++}$ 
in Ref. \cite{PABFF} decreases from 3.8 mb at 1 GeV to 2.9 mb at 1.38 GeV.
Elastic phase shifts for $K^+ K^-$ scattering were extracted from
$\pi^- p \to K^+ K^- n$ and $\pi^+ n \to K^+ K^- p$ in Ref. \cite{CADKP}.
A variety of theoretical approaches 
\cite{Roy,GL,DHT,Hannah,GO,ZB,DP,JPHS,OO1,BBO,SYZ,NA,OO2,BS} apply to
elastic meson-meson scattering.

Experimental data on $\pi \pi \to \rho$ are given 
in Ref. \cite{FMMR} and the cross section for $\pi K \to K^*$ in vacuum was 
estimated in Ref. \cite{Ko}. 
On the basis of the process that a quark in one initial 
meson and an antiquark in the other initial meson annihilate into a gluon and 
subsequently the gluon is absorbed by the other antiquark or quark,
2-to-1 meson-meson scattering has been studied in Ref. \cite{YXW}, 
and the resulting cross sections
for $\pi \pi \to \rho$ and $\pi K \to K^*$ agree with the empirical data given
in Refs. \cite{FMMR,Ko}. Cross sections for the reactions
$\pi^- \pi^- \to \pi^- \pi^- \pi^+ \pi^-$ and $\pi^- \pi^- \to \pi^- \pi^- 
\pi^0 \pi^0$ were measured in Refs. \cite{CFSW,LCFMP}, and the cross section
goes up when the dipion mass increases from 0.8 GeV.
The cross section for $\pi^- K^- \to \pi^- \pi^- K^0 + \pi^0 \pi^- K^-$ was 
investigated in Ref. \cite{JMTVH}. 
In the present work we study 2-to-3 meson-meson 
scattering based on the process that a gluon is created from a quark or an 
antiquark
in the two initial mesons, and the gluon then creates a quark and an antiquark.
We note that 2-to-3 meson-meson scattering has not yet been studied in theory.

Inelastic 2-to-2 meson-meson scattering has been studied in Refs. 
\cite{LDHS,BKWX,LX,SXW,PR}. The reactions 
$\pi \pi \to K \bar K$, $\rho \rho \to K \bar K$, $\pi \rho \to K \bar {K}^*$, 
and $\pi \rho \to K^* \bar {K}$ can be described by effective 
meson Lagrangians. The cross sections for the four reactions have been obtained
from the exchange of either a
kaon or a vector kaon between the two colliding mesons  \cite{LDHS,BKWX}. 
A study of $\pi \pi \to K\bar K$
scattering by means of partial-wave dispersion relations of the Roy-Steiner
type is performed in Ref. \cite{PR}, and precise parametrizations of 
the $S$, $P$, and $D$ partial waves in the $\pi\pi \to K\bar K$ scattering 
amplitude are obtained from the data.
To generate all the resonances with isospin 0 and masses below 2 GeV,
$S$-wave meson-meson scattering for total isospin $I=0$ and 1/2 is
studied in Ref. \cite{AO} with 13 coupled channels. Their $S$-wave
phase shifts and modulus for $\pi \pi \to K\bar K$ for $I=0$ agree with 
experimental data, and $\pi \pi \to \rho \rho$ is determined by minimal 
coupling. 
In terms of quark degrees of freedom some reactions 
are mainly governed by quark interchange, quark-antiquark annihilation and 
creation, or both. For example, $\pi\pi \to \rho\rho$ for total isospin $I=2$ 
and $\pi K \to \rho K^*$ for $I=3/2$ involve quark interchange \cite{LX}; 
$\pi \pi \to \rho \rho$ for $I=1$ and $\pi \rho \to K \bar {K}^\ast$
involve quark-antiquark annihilation and creation \cite{SXW}; 
$\pi \pi \to \rho \rho$ for $I=0$ and $\pi K \to \rho K^*$ for $I=1/2$ 
involve quark interchange as well as quark-antiquark annihilation and 
creation \cite{YXW,SXW}. The reactions governed by quark interchange as well
as quark-antiquark annihilation and creation have the characteristic feature 
that 
close to threshold quark interchange dominates the reactions near the critical 
temperature, and in the other energy region quark-antiquark annihilation 
and creation may dominate the reactions.

In hadronic matter created in relativistic heavy-ion
collisions at the Relativistic Heavy Ion Collider and at the Large Hadron 
Collider, thermal equilibrium is established by elastic meson-meson
scattering. Since inelastic meson-meson scattering alters the meson number,
chemical equilibrium is determined by inelastic meson-meson scattering. 
Exactly how thermal equilibrium is established and how chemical equilibrium is 
established are two important issues of
hadronic matter. In lead-lead collisions and in xenon-xenon collisions 
at the Large Hadron Collider,
meson momentum measured by the ATLAS Collaboration, the CMS Collaboration,
and the ALICE Collaboration goes up to 1000 GeV/$c$ 
\cite{ATLAS1000,CMS1000,ALICE1000}. A meson of such large momenta 
in collision with another meson in hadronic
matter may yield three or more mesons. Two-to-three meson-meson scattering
affects chemical equilibrium. Therefore, we need to study the 2-to-3
meson-meson scattering in hadronic matter.

This paper is organized as follows. In Sect.~II we present eight Feynman
diagrams for 2-to-3 meson-meson scattering, the transition amplitudes
corresponding to the eight diagrams, and cross sections related to the 
transition amplitudes. In Sect.~III we derive a transition potential for the
process that a gluon is created from a quark or an antiquark and the gluon
creates a quark and an antiquark. In Sect.~IV we 
calculate color, spin, and flavor matrix elements in the transition amplitudes.
In Sect.~V numerical cross sections are presented 
and relevant discussions are given. In Sect.~VI we summarize the present work.

\vspace{0.5cm}
\leftline{\bf II. CROSS-SECTION FORMULAS}
\vspace{0.5cm}

Meson $A$ contains quark $q_1$ and antiquark $\bar{q}_1$, and meson $B$ has 
quark $q_2$ and antiquark $\bar{q}_2$. In the collision of mesons $A$ and $B$ 
a constituent quark or antiquark may emit a virtual gluon which subsequently 
splits into quark $q_3$ and antiquark $\bar{q}_4$. The three quarks and 
antiquarks then 
combine into mesons $C_1$, $C_2$, and $C_3$. Four Feynman diagrams are shown 
in Fig. 1 
for $A(q_1\bar{q}_1)+B(q_2\bar{q}_2) \to C_1(q_1\bar{q}_4)+C_2(q_2\bar{q}_1)
+C_3(q_3\bar{q}_2)$, and four other diagrams
in Fig. 2 for $A(q_1\bar{q}_1)+B(q_2\bar{q}_2) \to 
C_1(q_1\bar{q}_2)+C_2(q_2\bar{q}_4)+C_3(q_3\bar{q}_1)$.
Diagram ${\rm D}_1$ (${\rm D}_2$, ${\rm D}_3$, ${\rm D}_4$) in Fig. 1 involves
the emission of a gluon from $q_1$ ($\bar{q}_1$, $q_2$, 
$\bar{q}_2$) and the subsequent splitting of the gluon into $q_3$ and 
$\bar{q}_4$, and diagram ${\rm D}_5$ (${\rm D}_6$, ${\rm D}_7$, ${\rm D}_8$) 
in Fig. 2 also involves this process.
Denote the energy of meson $A$ ($B$, $C_1$, $C_2$, $C_3$) by $E_A$ ($E_B$, 
$E_{C_1}$, $E_{C_2}$, $E_{C_3}$). The total energy of the two initial mesons
is $E_{\rm i}=E_A+E_B$, and the total energy of the three final mesons is 
$E_{\rm f}=E_{C_1}+E_{C_2}+E_{C_3}$.
The $S$-matrix element for $A+B \to C_1+C_2+C_3$ is
\begin{eqnarray}
S_{\rm fi} & = & \delta_{\rm fi} - 2\pi i \delta (E_{\rm f} - E_{\rm i})
(<C_1,C_2,C_3 \mid V_{{\rm D}_1} \mid A,B> 
+ <C_1,C_2,C_3 \mid V_{{\rm D}_2} \mid A,B> \nonumber\\
& &
+ <C_1,C_2,C_3 \mid V_{{\rm D}_3} \mid A,B> 
+ <C_1,C_2,C_3 \mid V_{{\rm D}_4} \mid A,B> \nonumber\\
& &
+ <C_1,C_2,C_3 \mid V_{{\rm D}_5} \mid A,B> 
+ <C_1,C_2,C_3 \mid V_{{\rm D}_6} \mid A,B> \nonumber\\
& &
+ <C_1,C_2,C_3 \mid V_{{\rm D}_7} \mid A,B> 
+ <C_1,C_2,C_3 \mid V_{{\rm D}_8} \mid A,B>) ,
\end{eqnarray}
where $V_{{\rm D}_1}$ ($V_{{\rm D}_2}$, $V_{{\rm D}_3}$, $V_{{\rm D}_4}$) 
represents the transition potential for $q_1 \to q_1+q_3+\bar{q}_4$
($\bar{q}_1 \to \bar{q}_1+q_3+\bar{q}_4$, $q_2 \to q_2+q_3+\bar{q}_4$,
$\bar{q}_2 \to \bar{q}_2+q_3+\bar{q}_4$) in diagram ${\rm D}_1$ (${\rm D}_2$, 
${\rm D}_3$, ${\rm D}_4$), and 
$V_{{\rm D}_5}$ ($V_{{\rm D}_6}$, $V_{{\rm D}_7}$, $V_{{\rm D}_8}$) 
represents the transition potential for $q_1 \to q_1+q_3+\bar{q}_4$
($\bar{q}_1 \to \bar{q}_1+q_3+\bar{q}_4$, $q_2 \to q_2+q_3+\bar{q}_4$,
$\bar{q}_2 \to \bar{q}_2+q_3+\bar{q}_4$) in diagram ${\rm D}_5$ (${\rm D}_6$, 
${\rm D}_7$, ${\rm D}_8$).
Let $\vec {P}_{ab}$, $\vec {R}_{ab}$, and $\vec {r}_{ab}$ be the total
momentum, the center-of-mass coordinate, and the relative coordinate of 
constituents $a$ and $b$, respectively. 
The wave function $\mid A,B>$ of mesons $A$ and $B$ is
\begin{equation}
\psi_{q_1\bar {q}_1, q_2\bar {q}_2}=
\frac {e^{i\vec {P}_{q_1\bar {q}_1}\cdot \vec {R}_{q_1\bar {q}_1}}}{\sqrt V}
\psi_{q_1\bar {q}_1} (\vec {r}_{q_1\bar {q}_1})
\frac {e^{i\vec {P}_{q_2\bar {q}_2}\cdot \vec {R}_{q_2\bar {q}_2}}}{\sqrt V}
\psi_{q_2\bar {q}_2} (\vec {r}_{q_2\bar {q}_2}).
\end{equation}
The wave function $\mid C_1,C_2,C_3>$ of mesons $C_1$, $C_2$, and $C_3$ is
\begin{equation}
\psi_{q_1\bar {q}_4,q_2\bar {q}_1,q_3\bar {q}_2}=
\frac {e^{i\vec {P}_{q_1\bar {q}_4}\cdot \vec {R}_{q_1\bar {q}_4}}}{\sqrt V}
\psi_{q_1\bar {q}_4} (\vec {r}_{q_1\bar {q}_4})
\frac {e^{i\vec {P}_{q_2\bar {q}_1}\cdot \vec {R}_{q_2\bar {q}_1}}}{\sqrt V}
\psi_{q_2\bar {q}_1} (\vec {r}_{q_2\bar {q}_1})
\frac {e^{i\vec {P}_{q_3\bar {q}_2}\cdot \vec {R}_{q_3\bar {q}_2}}}{\sqrt V}
\psi_{q_3\bar {q}_2} (\vec {r}_{q_3\bar {q}_2}),
\end{equation}
corresponding to the four diagrams in Fig. 1 or
\begin{equation}
\psi_{q_1\bar {q}_2,q_2\bar {q}_4,q_3\bar {q}_1}=
\frac {e^{i\vec {P}_{q_1\bar {q}_2}\cdot \vec {R}_{q_1\bar {q}_2}}}{\sqrt V}
\psi_{q_1\bar {q}_2} (\vec {r}_{q_1\bar {q}_2})
\frac {e^{i\vec {P}_{q_2\bar {q}_4}\cdot \vec {R}_{q_2\bar {q}_4}}}{\sqrt V}
\psi_{q_2\bar {q}_4} (\vec {r}_{q_2\bar {q}_4})
\frac {e^{i\vec {P}_{q_3\bar {q}_1}\cdot \vec {R}_{q_3\bar {q}_1}}}{\sqrt V}
\psi_{q_3\bar {q}_1} (\vec {r}_{q_3\bar {q}_1}),
\end{equation}
corresponding to the four diagrams in Fig. 2. 
The mesonic quark-antiquark wave function $\psi_{ab}(\vec{r}_{ab})$ is the 
product of the color wave function, the spin wave function, 
the flavor wave function, and the relative-motion wave function of 
constituents $a$ and $b$. Every meson wave 
function is normalized in the volume $V$.

From the $S$-matrix element we derive the transition amplitudes corresponding 
to the eight Feynman diagrams in Figs. 1 and 2. From the transition amplitudes
we obtain the unpolarized cross section for $A+B \to C_1+C_2+C_3$.
The position vector and the mass of constituent $c$ are denoted by $\vec{r}_c$
and $m_c$, respectively.

We first consider the four diagrams in Fig. 1. The five independent 
constituent position-vectors are $\vec{r}_{q_1}$, $\vec{r}_{\bar{q}_1}$, 
$\vec{r}_{q_2}$, $\vec{r}_{\bar{q}_2}$, and $\vec{r}_{q_3}$. 
They are related to the relative coordinates ($\vec {r}_{q_1\bar{q}_1}$, 
$\vec {r}_{q_2\bar{q}_2}$) and the center-of-mass coordinates 
($\vec{R}_{q_1\bar{q}_4}$, $\vec{R}_{q_2\bar{q}_1}$, 
$\vec{R}_{q_3\bar{q}_2}$) by
\begin{eqnarray}
\vec{r}_{q_1} & = & \frac{m_{\bar{q}_1}m_{\bar{q}_2}m_{\bar{q}_4}}
{m_{q_1}m_{q_2}m_{q_3}+m_{\bar{q}_1}m_{\bar{q}_2}m_{\bar{q}_4}}
\vec {r}_{q_1\bar{q}_1}
-\frac{m_{q_2}m_{\bar{q}_2}m_{\bar{q}_4}}
{m_{q_1}m_{q_2}m_{q_3}+m_{\bar{q}_1}m_{\bar{q}_2}m_{\bar{q}_4}}
\vec {r}_{q_2\bar{q}_2}
          \nonumber   \\ 
& + & \frac{m_{q_2}m_{q_3}(m_{q_1}+m_{\bar{q}_4})}
{m_{q_1}m_{q_2}m_{q_3}+m_{\bar{q}_1}m_{\bar{q}_2}m_{\bar{q}_4}}
\vec{R}_{q_1\bar{q}_4}
+\frac{m_{\bar{q}_2}m_{\bar{q}_4}(m_{q_2}+m_{\bar{q}_1})}
{m_{q_1}m_{q_2}m_{q_3}+m_{\bar{q}_1}m_{\bar{q}_2}m_{\bar{q}_4}}
\vec{R}_{q_2\bar{q}_1}
          \nonumber   \\ 
& - & \frac{m_{q_2}m_{\bar{q}_4}(m_{q_3}+m_{\bar{q}_2})}
{m_{q_1}m_{q_2}m_{q_3}+m_{\bar{q}_1}m_{\bar{q}_2}m_{\bar{q}_4}}
\vec{R}_{q_3\bar{q}_2},
\end{eqnarray}
\begin{eqnarray}
\vec{r}_{\bar{q}_1} & = & -\frac{m_{q_1}m_{q_2}m_{q_3}}
{m_{q_1}m_{q_2}m_{q_3}+m_{\bar{q}_1}m_{\bar{q}_2}m_{\bar{q}_4}}
\vec {r}_{q_1\bar{q}_1}
-\frac{m_{q_2}m_{\bar{q}_2}m_{\bar{q}_4}}
{m_{q_1}m_{q_2}m_{q_3}+m_{\bar{q}_1}m_{\bar{q}_2}m_{\bar{q}_4}}
\vec {r}_{q_2\bar{q}_2}
          \nonumber   \\ 
& + & \frac{m_{q_2}m_{q_3}(m_{q_1}+m_{\bar{q}_4})}
{m_{q_1}m_{q_2}m_{q_3}+m_{\bar{q}_1}m_{\bar{q}_2}m_{\bar{q}_4}}
\vec{R}_{q_1\bar{q}_4}
+ \frac{m_{\bar{q}_2}m_{\bar{q}_4}(m_{q_2}+m_{\bar{q}_1})}
{m_{q_1}m_{q_2}m_{q_3}+m_{\bar{q}_1}m_{\bar{q}_2}m_{\bar{q}_4}}
\vec{R}_{q_2\bar{q}_1}
          \nonumber   \\ 
& - & \frac{m_{q_2}m_{\bar{q}_4}(m_{q_3}+m_{\bar{q}_2})}
{m_{q_1}m_{q_2}m_{q_3}+m_{\bar{q}_1}m_{\bar{q}_2}m_{\bar{q}_4}}
\vec{R}_{q_3\bar{q}_2},
\end{eqnarray}
\begin{eqnarray}
\vec{r}_{q_2} & = & \frac{m_{q_1}m_{q_3}m_{\bar{q}_1}}
{m_{q_1}m_{q_2}m_{q_3}+m_{\bar{q}_1}m_{\bar{q}_2}m_{\bar{q}_4}}
\vec {r}_{q_1\bar{q}_1}
+\frac{m_{\bar{q}_1}m_{\bar{q}_2}m_{\bar{q}_4}}
{m_{q_1}m_{q_2}m_{q_3}+m_{\bar{q}_1}m_{\bar{q}_2}m_{\bar{q}_4}}
\vec {r}_{q_2\bar{q}_2}
          \nonumber   \\ 
& - & \frac{m_{q_3}m_{\bar{q}_1}(m_{q_1}+m_{\bar{q}_4})}
{m_{q_1}m_{q_2}m_{q_3}+m_{\bar{q}_1}m_{\bar{q}_2}m_{\bar{q}_4}}
\vec{R}_{q_1\bar{q}_4}
+ \frac{m_{q_1}m_{q_3}(m_{q_2}+m_{\bar{q}_1})}
{m_{q_1}m_{q_2}m_{q_3}+m_{\bar{q}_1}m_{\bar{q}_2}m_{\bar{q}_4}}
\vec{R}_{q_2\bar{q}_1}
          \nonumber   \\ 
& + & \frac{m_{\bar{q}_1}m_{\bar{q}_4}(m_{q_3}+m_{\bar{q}_2})}
{m_{q_1}m_{q_2}m_{q_3}+m_{\bar{q}_1}m_{\bar{q}_2}m_{\bar{q}_4}}
\vec{R}_{q_3\bar{q}_2},
\end{eqnarray}
\begin{eqnarray}
\vec{r}_{\bar{q}_2} & = & \frac{m_{q_1}m_{q_3}m_{\bar{q}_1}}
{m_{q_1}m_{q_2}m_{q_3}+m_{\bar{q}_1}m_{\bar{q}_2}m_{\bar{q}_4}}
\vec {r}_{q_1\bar{q}_1}
-\frac{m_{q_1}m_{q_2}m_{q_3}}
{m_{q_1}m_{q_2}m_{q_3}+m_{\bar{q}_1}m_{\bar{q}_2}m_{\bar{q}_4}}
\vec {r}_{q_2\bar{q}_2}
          \nonumber   \\ 
& - & \frac{m_{q_3}m_{\bar{q}_1}(m_{q_1}+m_{\bar{q}_4})}
{m_{q_1}m_{q_2}m_{q_3}+m_{\bar{q}_1}m_{\bar{q}_2}m_{\bar{q}_4}}
\vec{R}_{q_1\bar{q}_4}
+ \frac{m_{q_1}m_{q_3}(m_{q_2}+m_{\bar{q}_1})}
{m_{q_1}m_{q_2}m_{q_3}+m_{\bar{q}_1}m_{\bar{q}_2}m_{\bar{q}_4}}
\vec{R}_{q_2\bar{q}_1}
          \nonumber   \\ 
& + & \frac{m_{\bar{q}_1}m_{\bar{q}_4}(m_{q_3}+m_{\bar{q}_2})}
{m_{q_1}m_{q_2}m_{q_3}+m_{\bar{q}_1}m_{\bar{q}_2}m_{\bar{q}_4}}
\vec{R}_{q_3\bar{q}_2},
\end{eqnarray}
\begin{eqnarray}
\vec{r}_{q_3} & = & -\frac{m_{q_1}m_{\bar{q}_1}m_{\bar{q}_2}}
{m_{q_1}m_{q_2}m_{q_3}+m_{\bar{q}_1}m_{\bar{q}_2}m_{\bar{q}_4}}
\vec {r}_{q_1\bar{q}_1}
+\frac{m_{q_1}m_{q_2}m_{\bar{q}_2}}
{m_{q_1}m_{q_2}m_{q_3}+m_{\bar{q}_1}m_{\bar{q}_2}m_{\bar{q}_4}}
\vec {r}_{q_2\bar{q}_2}
          \nonumber   \\ 
& + & \frac{m_{\bar{q}_1}m_{\bar{q}_2}(m_{q_1}+m_{\bar{q}_4})}
{m_{q_1}m_{q_2}m_{q_3}+m_{\bar{q}_1}m_{\bar{q}_2}m_{\bar{q}_4}}
\vec{R}_{q_1\bar{q}_4}
- \frac{m_{q_1}m_{\bar{q}_2}(m_{q_2}+m_{\bar{q}_1})}
{m_{q_1}m_{q_2}m_{q_3}+m_{\bar{q}_1}m_{\bar{q}_2}m_{\bar{q}_4}}
\vec{R}_{q_2\bar{q}_1}
          \nonumber   \\ 
& + & \frac{m_{q_1}m_{q_2}(m_{q_3}+m_{\bar{q}_2})}
{m_{q_1}m_{q_2}m_{q_3}+m_{\bar{q}_1}m_{\bar{q}_2}m_{\bar{q}_4}}
\vec{R}_{q_3\bar{q}_2},
\end{eqnarray}
which lead to
\begin{eqnarray}
& & d\vec{r}_{q_1} d\vec{r}_{\bar{q}_1} d\vec{r}_{q_2} d\vec{r}_{\bar{q}_2}
d\vec{r}_{q_3} 
     \nonumber   \\
& = & \frac{(m_{q_1}+m_{\bar{q}_4})^3(m_{q_2}+m_{\bar{q}_1})^3
(m_{q_3}+m_{\bar{q}_2})^3}
{(m_{q_1}m_{q_2}m_{q_3}+m_{\bar{q}_1}m_{\bar{q}_2}m_{\bar{q}_4})^3}
d\vec{r}_{q_1\bar{q}_1} d\vec{r}_{q_2\bar{q}_2} d\vec{R}_{q_1\bar{q}_4} 
d\vec{R}_{q_2\bar{q}_1} d\vec{R}_{q_3\bar{q}_2}.
\end{eqnarray}

Let $\vec {R}_{\rm total}$ be the center-of-mass coordinate of the
initial or final mesons. 
Denote the three-dimensional momentum of meson $A$ ($B$, $C_1$, $C_2$, $C_3$)
by $\vec{P}_A$ ($\vec{P}_B$, $\vec{P}_{C_1}$, $\vec{P}_{C_2}$, 
$\vec{P}_{C_3}$). The total momentum of the two initial
mesons is $\vec{P}_{\rm i}=\vec{P}_A+\vec{P}_B$, and
the total momentum of the three final mesons is 
$\vec{P}_{\rm f}=\vec{P}_{C_1}+\vec{P}_{C_2}+\vec{P}_{C_3}$.
From the transition potential $V_{\rm D_1}$ and the wave functions 
of the initial and final mesons, we have for diagram ${\rm D}_1$:
\begin{eqnarray}
& & <C_1, C_2, C_3 \mid V_{{\rm D}_1} \mid A, B> 
    \nonumber    \\
& = & <q_1\bar{q}_4, q_2\bar{q}_1, q_3\bar{q}_2 \mid V_{{\rm D}_1} 
\mid q_1\bar{q}_1, q_2\bar{q}_2>
    \nonumber    \\
& = & \int d\vec{r}_{q_1} d\vec{r}_{\bar{q}_1} d\vec{r}_{q_2} 
d\vec{r}_{\bar{q}_2} d\vec{r}_{q_3} 
\frac {e^{-i\vec{P}_{q_1\bar{q}_4}\cdot\vec{R}_{q_1\bar{q}_4}}}{\sqrt V}
\psi_{q_1\bar{q}_4}^+ (\vec{r}_{q_1\bar{q}_4})
\frac {e^{-i\vec{P}_{q_2\bar{q}_1}\cdot\vec{R}_{q_2\bar{q}_1}}}{\sqrt V}
\psi_{q_2\bar{q}_1}^+ (\vec{r}_{q_2\bar{q}_1})
    \nonumber    \\
& & \frac {e^{-i\vec{P}_{q_3\bar{q}_2}\cdot\vec{R}_{q_3\bar{q}_2}}}{\sqrt V}
\psi_{q_3\bar{q}_2}^+ (\vec{r}_{q_3\bar{q}_2})
V_{{\rm D}_1}
\frac {e^{i\vec{P}_{q_1\bar{q}_1}\cdot\vec{R}_{q_1\bar{q}_1}}}{\sqrt V}
\psi_{q_1\bar{q}_1}(\vec{r}_{q_1\bar{q}_1})
\frac {e^{i\vec{P}_{q_2\bar{q}_2}\cdot\vec{R}_{q_2\bar{q}_2}}}{\sqrt V}
\psi_{q_2\bar{q}_2}(\vec{r}_{q_2\bar{q}_2})
    \nonumber    \\
& = & \frac{(m_{q_1}+m_{\bar{q}_4})^3(m_{q_2}+m_{\bar{q}_1})^3
(m_{q_3}+m_{\bar{q}_2})^3}
{\sqrt {V^5}(m_{q_1}m_{q_2}m_{q_3}+m_{\bar{q}_1}m_{\bar{q}_2}m_{\bar{q}_4})^3}
\int d\vec{r}_{q_1\bar{q}_1} d\vec{r}_{q_2\bar{q}_2} d\vec{R}_{q_1\bar{q}_4}
d\vec{R}_{q_2\bar{q}_1} d\vec{R}_{q_3\bar{q}_2} 
    \nonumber    \\
& & \psi_{q_1\bar{q}_4}^+ (\vec{r}_{q_1\bar{q}_4})
\psi_{q_2\bar{q}_1}^+ (\vec{r}_{q_2\bar{q}_1})
\psi_{q_3\bar{q}_2}^+ (\vec{r}_{q_3\bar{q}_2})
V_{{\rm D}_1}
\psi_{q_1\bar{q}_1}(\vec{r}_{q_1\bar{q}_1})
\psi_{q_2\bar{q}_2}(\vec{r}_{q_2\bar{q}_2})
    \nonumber    \\
& & \exp (-i\vec{P}_{q_1\bar{q}_4} \cdot \vec{R}_{q_1\bar{q}_4}
-i\vec{P}_{q_2\bar{q}_1} \cdot \vec{R}_{q_2\bar{q}_1}
-i\vec{P}_{q_3\bar{q}_2} \cdot \vec{R}_{q_3\bar{q}_2}
+i\vec{P}_{q_1\bar{q}_1} \cdot \vec{R}_{q_1\bar{q}_1}
+i\vec{P}_{q_2\bar{q}_2} \cdot \vec{R}_{q_2\bar{q}_2}),
    \nonumber    \\
\end{eqnarray}
where $\psi_{ab}^+$ is the Hermitean conjugate of $\psi_{ab}$.
In the present work we limit ourselves to the case that at least two of the
three final mesons have the same mass. We thus define the variable 
$\vec{\rho}_X$ from the position vectors of the two mesons with equal masses
and the variable $\vec{\lambda}_X$ from the other meson. For example, supposing
that mesons $C_1(q_1\bar{q}_4)$ and $C_2(q_2\bar{q}_1)$ have equal masses, 
we define
\begin{equation}
\vec{\rho}_X=\frac{1}{\sqrt 2} (\vec{R}_{q_1\bar{q}_4}-\vec{R}_{q_2\bar{q}_1}),
\end{equation}
\begin{equation}
\vec{\lambda}_X=\frac{1}{\sqrt 6} (\vec{R}_{q_1\bar{q}_4}
+\vec{R}_{q_2\bar{q}_1}-2\vec{R}_{q_3\bar{q}_2}),
\end{equation}
which leads to 
\begin{equation}
d\vec{R}_{q_1\bar{q}_4}d\vec{R}_{q_2\bar{q}_1}d\vec{R}_{q_3\bar{q}_2}
=3\sqrt{3}d\vec{R}_{\rm total}d\vec{\rho}_Xd\vec{\lambda}_X.
\end{equation}
From the mass $m_{C_1}$ of meson $C_1$, the mass $m_{C_2}$ of meson $C_2$,
and the mass $m_{C_3}$ of meson $C_3$, we define
\begin{equation}
m_\rho=m_{C_1}=m_{C_2},
\end{equation}
\begin{equation}
m_\lambda=\frac{3m_{C_1}m_{C_3}}{2m_{C_1}+m_{C_3}}.
\end{equation}
Let $\vec{p}_{\rho_X}$ ($\vec{p}_{\lambda_X}$) be $m_\rho$ ($m_\lambda$) times
the derivative of $\vec{\rho}_X$ ($\vec{\lambda}_X$) with respect to time.
We then get
\begin{equation}
\vec{P}_{q_1\bar{q}_4} \cdot \vec{R}_{q_1\bar{q}_4}
+\vec{P}_{q_2\bar{q}_1} \cdot \vec{R}_{q_2\bar{q}_1}
+\vec{P}_{q_3\bar{q}_2} \cdot \vec{R}_{q_3\bar{q}_2}
=\vec{P}_{\rm f} \cdot \vec{R}_{\rm total}
+\vec{p}_{\rho_X} \cdot \vec{\rho}_X+\vec{p}_{\lambda_X} \cdot \vec{\lambda}_X.
\end{equation}
In terms of $\vec{\rho}_X$, $\vec{\lambda}_X$, $\vec{p}_{\rho_X}$, and 
$\vec{p}_{\lambda_X}$, we have
\begin{eqnarray}
& & <C_1, C_2, C_3 \mid V_{{\rm D}_1} \mid A, B> 
    \nonumber    \\
& = & \frac{3\sqrt{3}(m_{q_1}+m_{\bar{q}_4})^3(m_{q_2}+m_{\bar{q}_1})^3
(m_{q_3}+m_{\bar{q}_2})^3}
{\sqrt {V^5}(m_{q_1}m_{q_2}m_{q_3}+m_{\bar{q}_1}m_{\bar{q}_2}m_{\bar{q}_4})^3}
\int d\vec{r}_{q_1\bar{q}_1} d\vec{r}_{q_2\bar{q}_2}
d\vec{R}_{\rm total} d\vec{\rho}_X d\vec{\lambda}_X
    \nonumber    \\
& & \psi_{q_1\bar{q}_4}^+ (\vec{r}_{q_1\bar{q}_4})
\psi_{q_2\bar{q}_1}^+ (\vec{r}_{q_2\bar{q}_1})
\psi_{q_3\bar{q}_2}^+ (\vec{r}_{q_3\bar{q}_2})
V_{{\rm D}_1}
\psi_{q_1\bar{q}_1}(\vec{r}_{q_1\bar{q}_1})
\psi_{q_2\bar{q}_2}(\vec{r}_{q_2\bar{q}_2})
    \nonumber    \\
& & \exp (-i\vec{P}_{\rm f} \cdot \vec{R}_{\rm total}
-i\vec{p}_{\rho_X} \cdot \vec{\rho}_X 
-i\vec{p}_{\lambda_X} \cdot \vec{\lambda}_X
+i \vec{P}_{\rm i} \cdot \vec{R}_{\rm total}
+i\vec{p}_{q_1\bar{q}_1,q_2\bar{q}_2}\cdot\vec{r}_{q_1\bar{q}_1,q_2\bar{q}_2})
    \nonumber    \\
& = & (2\pi)^3 \delta^3 (\vec{P}_{\rm i} - \vec{P}_{\rm f}) 
\frac {{\cal M}_{{\rm D}_1} }
{\sqrt {V^5}\sqrt{2E_A2E_B2E_{C_1}2E_{C_2}2E_{C_3}}},
\end{eqnarray}
where $\vec {r}_{q_1\bar{q}_1,q_2\bar{q}_2}$ and 
$\vec {p}_{q_1\bar{q}_1,q_2\bar{q}_2}$
are the relative coordinate and the relative momentum of $q_1\bar {q}_1$ and 
$q_2\bar {q}_2$, respectively. ${\cal M}_{\rm D_1}$
is the transition amplitude given by
\begin{eqnarray}
{\cal M}_{{\rm D}_1}
& = & \sqrt {2E_A2E_B2E_{C_1}2E_{C_2}2E_{C_3}}
\frac{3\sqrt{3}(m_{q_1}+m_{\bar{q}_4})^3(m_{q_2}+m_{\bar{q}_1})^3
(m_{q_3}+m_{\bar{q}_2})^3}
{(m_{q_1}m_{q_2}m_{q_3}+m_{\bar{q}_1}m_{\bar{q}_2}m_{\bar{q}_4})^3}
    \nonumber    \\
& & \int d\vec{r}_{q_1\bar{q}_1}d\vec{r}_{q_2\bar{q}_2}
d\vec{\rho}_X d\vec{\lambda}_X
\psi_{q_1\bar{q}_4}^+ (\vec{r}_{q_1\bar{q}_4})
\psi_{q_2\bar{q}_1}^+ (\vec{r}_{q_2\bar{q}_1})
\psi_{q_3\bar{q}_2}^+ (\vec{r}_{q_3\bar{q}_2}) V_{{\rm D}_1}
    \nonumber    \\
& & \psi_{q_1\bar{q}_1}(\vec{r}_{q_1\bar{q}_1})
\psi_{q_2\bar{q}_2}(\vec{r}_{q_2\bar{q}_2})
\exp (-i\vec{p}_{\rho_X} \cdot \vec{\rho}_X 
-i\vec{p}_{\lambda_X} \cdot \vec{\lambda}_X 
+i\vec{p}_{q_1\bar{q}_1,q_2\bar{q}_2}\cdot\vec{r}_{q_1\bar{q}_1,q_2\bar{q}_2}).
    \nonumber    \\
\end{eqnarray}

For diagram ${\rm D}_2$ we have
\begin{eqnarray}
<C_1, C_2, C_3 \mid V_{{\rm D}_2}\mid A,B>
& = & <q_1\bar{q}_4, q_2\bar{q}_1, q_3\bar{q}_2 \mid 
V_{{\rm D}_2} \mid q_1\bar{q}_1, q_2\bar{q}_2>
    \nonumber    \\
& = & (2\pi)^3 \delta^3 (\vec{P}_{\rm i}-\vec{P}_{\rm f}) 
\frac {{\cal M}_{{\rm D}_2} }
{\sqrt {V^5}\sqrt{2E_A2E_B2E_{C_1}2E_{C_2}2E_{C_3}}},
\end{eqnarray}
where the transition amplitude ${\cal M}_{{\rm D}_2}$ is 
obtained from Eq. (19) by replacing $V_{{\rm D}_1}$ with 
$V_{{\rm D}_2}$. For diagram ${\rm D}_3$ we have
\begin{eqnarray}
<C_1, C_2, C_3 \mid V_{{\rm D}_3}\mid A,B>
& = & <q_1\bar{q}_4, q_2\bar{q}_1, q_3\bar{q}_2 \mid 
V_{{\rm D}_3} \mid q_1\bar{q}_1, q_2\bar{q}_2>
    \nonumber    \\
& = & (2\pi)^3 \delta^3 (\vec{P}_{\rm i}-\vec{P}_{\rm f}) 
\frac {{\cal M}_{{\rm D}_3} }
{\sqrt {V^5}\sqrt{2E_A2E_B2E_{C_1}2E_{C_2}2E_{C_3}}},
\end{eqnarray}
where the transition amplitude ${\cal M}_{{\rm D}_3}$ is 
obtained from Eq. (19) by replacing $V_{{\rm D}_1}$ with 
$V_{{\rm D}_3}$. For diagram ${\rm D}_4$ we have
\begin{eqnarray}
<C_1, C_2, C_3 \mid V_{{\rm D}_4}\mid A,B>
& = & <q_1\bar{q}_4, q_2\bar{q}_1, q_3\bar{q}_2 \mid 
V_{{\rm D}_4} \mid q_1\bar{q}_1, q_2\bar{q}_2>
    \nonumber    \\
& = & (2\pi)^3 \delta^3 (\vec{P}_{\rm i}-\vec{P}_{\rm f}) 
\frac {{\cal M}_{{\rm D}_4} }
{\sqrt {V^5}\sqrt{2E_A2E_B2E_{C_1}2E_{C_2}2E_{C_3}}},
\end{eqnarray}
where the transition amplitude ${\cal M}_{{\rm D}_4}$ is 
obtained from Eq. (19) by replacing $V_{{\rm D}_1}$ with $V_{{\rm D}_4}$.

Next, we consider the four diagrams in Fig. 2. The five independent
constituent position-vectors, $\vec{r}_{q_1}$, $\vec{r}_{\bar{q}_1}$, 
$\vec{r}_{q_2}$, $\vec{r}_{\bar{q}_2}$, and $\vec{r}_{q_3}$,
are related to $\vec {r}_{q_1\bar{q}_1}$, $\vec {r}_{q_2\bar{q}_2}$, 
$\vec{R}_{q_1\bar{q}_2}$, $\vec{R}_{q_2\bar{q}_4}$, and 
$\vec{R}_{q_3\bar{q}_1}$ by
\begin{eqnarray}
\vec{r}_{q_1} & = & \frac{m_{\bar{q}_1}m_{\bar{q}_2}m_{\bar{q}_4}}
{m_{q_1}m_{q_2}m_{q_3}+m_{\bar{q}_1}m_{\bar{q}_2}m_{\bar{q}_4}}
\vec {r}_{q_1\bar{q}_1}
+\frac{m_{q_2}m_{q_3}m_{\bar{q}_2}}
{m_{q_1}m_{q_2}m_{q_3}+m_{\bar{q}_1}m_{\bar{q}_2}m_{\bar{q}_4}}
\vec {r}_{q_2\bar{q}_2}
          \nonumber   \\ 
& + & \frac{m_{q_2}m_{q_3}(m_{q_1}+m_{\bar{q}_2})}
{m_{q_1}m_{q_2}m_{q_3}+m_{\bar{q}_1}m_{\bar{q}_2}m_{\bar{q}_4}}
\vec{R}_{q_1\bar{q}_2}
-\frac{m_{q_3}m_{\bar{q}_2}(m_{q_2}+m_{\bar{q}_4})}
{m_{q_1}m_{q_2}m_{q_3}+m_{\bar{q}_1}m_{\bar{q}_2}m_{\bar{q}_4}}
\vec{R}_{q_2\bar{q}_4}
          \nonumber   \\ 
& + & \frac{m_{\bar{q}_2}m_{\bar{q}_4}(m_{q_3}+m_{\bar{q}_1})}
{m_{q_1}m_{q_2}m_{q_3}+m_{\bar{q}_1}m_{\bar{q}_2}m_{\bar{q}_4}}
\vec{R}_{q_3\bar{q}_1},
\end{eqnarray}
\begin{eqnarray}
\vec{r}_{\bar{q}_1} & = & -\frac{m_{q_1}m_{q_2}m_{q_3}}
{m_{q_1}m_{q_2}m_{q_3}+m_{\bar{q}_1}m_{\bar{q}_2}m_{\bar{q}_4}}
\vec {r}_{q_1\bar{q}_1}
+\frac{m_{q_2}m_{q_3}m_{\bar{q}_2}}
{m_{q_1}m_{q_2}m_{q_3}+m_{\bar{q}_1}m_{\bar{q}_2}m_{\bar{q}_4}}
\vec {r}_{q_2\bar{q}_2}
          \nonumber   \\ 
& + & \frac{m_{q_2}m_{q_3}(m_{q_1}+m_{\bar{q}_2})}
{m_{q_1}m_{q_2}m_{q_3}+m_{\bar{q}_1}m_{\bar{q}_2}m_{\bar{q}_4}}
\vec{R}_{q_1\bar{q}_2}
- \frac{m_{q_3}m_{\bar{q}_2}(m_{q_2}+m_{\bar{q}_4})}
{m_{q_1}m_{q_2}m_{q_3}+m_{\bar{q}_1}m_{\bar{q}_2}m_{\bar{q}_4}}
\vec{R}_{q_2\bar{q}_4}
          \nonumber   \\ 
& + & \frac{m_{\bar{q}_2}m_{\bar{q}_4}(m_{q_3}+m_{\bar{q}_1})}
{m_{q_1}m_{q_2}m_{q_3}+m_{\bar{q}_1}m_{\bar{q}_2}m_{\bar{q}_4}}
\vec{R}_{q_3\bar{q}_1},
\end{eqnarray}
\begin{eqnarray}
\vec{r}_{q_2} & = & -\frac{m_{q_1}m_{\bar{q}_1}m_{\bar{q}_4}}
{m_{q_1}m_{q_2}m_{q_3}+m_{\bar{q}_1}m_{\bar{q}_2}m_{\bar{q}_4}}
\vec {r}_{q_1\bar{q}_1}
+\frac{m_{\bar{q}_1}m_{\bar{q}_2}m_{\bar{q}_4}}
{m_{q_1}m_{q_2}m_{q_3}+m_{\bar{q}_1}m_{\bar{q}_2}m_{\bar{q}_4}}
\vec {r}_{q_2\bar{q}_2}
          \nonumber   \\ 
& + & \frac{m_{\bar{q}_1}m_{\bar{q}_4}(m_{q_1}+m_{\bar{q}_2})}
{m_{q_1}m_{q_2}m_{q_3}+m_{\bar{q}_1}m_{\bar{q}_2}m_{\bar{q}_4}}
\vec{R}_{q_1\bar{q}_2}
+ \frac{m_{q_1}m_{q_3}(m_{q_2}+m_{\bar{q}_4})}
{m_{q_1}m_{q_2}m_{q_3}+m_{\bar{q}_1}m_{\bar{q}_2}m_{\bar{q}_4}}
\vec{R}_{q_2\bar{q}_4}
          \nonumber   \\ 
& - & \frac{m_{q_1}m_{\bar{q}_4}(m_{q_3}+m_{\bar{q}_1})}
{m_{q_1}m_{q_2}m_{q_3}+m_{\bar{q}_1}m_{\bar{q}_2}m_{\bar{q}_4}}
\vec{R}_{q_3\bar{q}_1},
\end{eqnarray}
\begin{eqnarray}
\vec{r}_{\bar{q}_2} & = & -\frac{m_{q_1}m_{\bar{q}_1}m_{\bar{q}_4}}
{m_{q_1}m_{q_2}m_{q_3}+m_{\bar{q}_1}m_{\bar{q}_2}m_{\bar{q}_4}}
\vec {r}_{q_1\bar{q}_1}
-\frac{m_{q_1}m_{q_2}m_{q_3}}
{m_{q_1}m_{q_2}m_{q_3}+m_{\bar{q}_1}m_{\bar{q}_2}m_{\bar{q}_4}}
\vec {r}_{q_2\bar{q}_2}
          \nonumber   \\ 
& + & \frac{m_{\bar{q}_1}m_{\bar{q}_4}(m_{q_1}+m_{\bar{q}_2})}
{m_{q_1}m_{q_2}m_{q_3}+m_{\bar{q}_1}m_{\bar{q}_2}m_{\bar{q}_4}}
\vec{R}_{q_1\bar{q}_2}
+ \frac{m_{q_1}m_{q_3}(m_{q_2}+m_{\bar{q}_4})}
{m_{q_1}m_{q_2}m_{q_3}+m_{\bar{q}_1}m_{\bar{q}_2}m_{\bar{q}_4}}
\vec{R}_{q_2\bar{q}_4}
          \nonumber   \\ 
& - & \frac{m_{q_1}m_{\bar{q}_4}(m_{q_3}+m_{\bar{q}_1})}
{m_{q_1}m_{q_2}m_{q_3}+m_{\bar{q}_1}m_{\bar{q}_2}m_{\bar{q}_4}}
\vec{R}_{q_3\bar{q}_1},
\end{eqnarray}
\begin{eqnarray}
\vec{r}_{q_3} & = & \frac{m_{q_1}m_{q_2}m_{\bar{q}_1}}
{m_{q_1}m_{q_2}m_{q_3}+m_{\bar{q}_1}m_{\bar{q}_2}m_{\bar{q}_4}}
\vec {r}_{q_1\bar{q}_1}
-\frac{m_{q_2}m_{\bar{q}_1}m_{\bar{q}_2}}
{m_{q_1}m_{q_2}m_{q_3}+m_{\bar{q}_1}m_{\bar{q}_2}m_{\bar{q}_4}}
\vec {r}_{q_2\bar{q}_2}
          \nonumber   \\ 
& - & \frac{m_{q_2}m_{\bar{q}_1}(m_{q_1}+m_{\bar{q}_2})}
{m_{q_1}m_{q_2}m_{q_3}+m_{\bar{q}_1}m_{\bar{q}_2}m_{\bar{q}_4}}
\vec{R}_{q_1\bar{q}_2}
+ \frac{m_{\bar{q}_1}m_{\bar{q}_2}(m_{q_2}+m_{\bar{q}_4})}
{m_{q_1}m_{q_2}m_{q_3}+m_{\bar{q}_1}m_{\bar{q}_2}m_{\bar{q}_4}}
\vec{R}_{q_2\bar{q}_4}
          \nonumber   \\ 
& + & \frac{m_{q_1}m_{q_2}(m_{q_3}+m_{\bar{q}_1})}
{m_{q_1}m_{q_2}m_{q_3}+m_{\bar{q}_1}m_{\bar{q}_2}m_{\bar{q}_4}}
\vec{R}_{q_3\bar{q}_1},
\end{eqnarray}
which lead to
\begin{eqnarray}
& & d\vec{r}_{q_1} d\vec{r}_{\bar{q}_1} d\vec{r}_{q_2} d\vec{r}_{\bar{q}_2}
d\vec{r}_{q_3}
        \nonumber  \\
& = & \frac{(m_{q_1}+m_{\bar{q}_2})^3(m_{q_2}+m_{\bar{q}_4})^3
(m_{q_3}+m_{\bar{q}_1})^3}
{(m_{q_1}m_{q_2}m_{q_3}+m_{\bar{q}_1}m_{\bar{q}_2}m_{\bar{q}_4})^3}
d\vec{r}_{q_1\bar{q}_1} d\vec{r}_{q_2\bar{q}_2} d\vec{R}_{q_1\bar{q}_2} 
d\vec{R}_{q_2\bar{q}_4} d\vec{R}_{q_3\bar{q}_1} .
\end{eqnarray}

From the transition potential $V_{\rm D_5}$ and
the wave functions of the initial and final mesons, 
we have for diagram ${\rm D}_5$ in Fig. 2:
\begin{eqnarray}
& & <C_1, C_2, C_3 \mid V_{{\rm D}_5} \mid A, B> 
    \nonumber    \\
& = & <q_1\bar{q}_2, q_2\bar{q}_4, q_3\bar{q}_1 \mid V_{{\rm D}_5} 
\mid q_1\bar{q}_1, q_2\bar{q}_2>
    \nonumber    \\
& = & \int d\vec{r}_{q_1} d\vec{r}_{\bar{q}_1} d\vec{r}_{q_2} 
d\vec{r}_{\bar{q}_2} d\vec{r}_{q_3} 
\frac {e^{-i\vec{P}_{q_1\bar{q}_2}\cdot\vec{R}_{q_1\bar{q}_2}}}{\sqrt V}
\psi_{q_1\bar{q}_2}^+ (\vec{r}_{q_1\bar{q}_2})
\frac {e^{-i\vec{P}_{q_2\bar{q}_4}\cdot\vec{R}_{q_2\bar{q}_4}}}{\sqrt V}
\psi_{q_2\bar{q}_4}^+ (\vec{r}_{q_2\bar{q}_4})
    \nonumber    \\
& & \frac {e^{-i\vec{P}_{q_3\bar{q}_1}\cdot\vec{R}_{q_3\bar{q}_1}}}{\sqrt V}
\psi_{q_3\bar{q}_1}^+ (\vec{r}_{q_3\bar{q}_1})
V_{{\rm D}_5}
\frac {e^{i\vec{P}_{q_1\bar{q}_1}\cdot\vec{R}_{q_1\bar{q}_1}}}{\sqrt V}
\psi_{q_1\bar{q}_1}(\vec{r}_{q_1\bar{q}_1})
\frac {e^{i\vec{P}_{q_2\bar{q}_2}\cdot\vec{R}_{q_2\bar{q}_2}}}{\sqrt V}
\psi_{q_2\bar{q}_2}(\vec{r}_{q_2\bar{q}_2})
    \nonumber    \\
& = & \frac{(m_{q_1}+m_{\bar{q}_2})^3(m_{q_2}+m_{\bar{q}_4})^3
(m_{q_3}+m_{\bar{q}_1})^3}
{\sqrt {V^5}(m_{q_1}m_{q_2}m_{q_3}+m_{\bar{q}_1}m_{\bar{q}_2}m_{\bar{q}_4})^3}
\int d\vec{r}_{q_1\bar{q}_1} d\vec{r}_{q_2\bar{q}_2} d\vec{R}_{q_1\bar{q}_2}
d\vec{R}_{q_2\bar{q}_4} d\vec{R}_{q_3\bar{q}_1} 
    \nonumber    \\
& & \psi_{q_1\bar{q}_2}^+ (\vec{r}_{q_1\bar{q}_2})
\psi_{q_2\bar{q}_4}^+ (\vec{r}_{q_2\bar{q}_4})
\psi_{q_3\bar{q}_1}^+ (\vec{r}_{q_3\bar{q}_1})
V_{{\rm D}_5}
\psi_{q_1\bar{q}_1}(\vec{r}_{q_1\bar{q}_1})
\psi_{q_2\bar{q}_2}(\vec{r}_{q_2\bar{q}_2})
    \nonumber    \\
& & \exp (-i\vec{P}_{q_1\bar{q}_2} \cdot \vec{R}_{q_1\bar{q}_2}
-i\vec{P}_{q_2\bar{q}_4} \cdot \vec{R}_{q_2\bar{q}_4}
-i\vec{P}_{q_3\bar{q}_1} \cdot \vec{R}_{q_3\bar{q}_1}
+i\vec{P}_{q_1\bar{q}_1} \cdot \vec{R}_{q_1\bar{q}_1}
+i\vec{P}_{q_2\bar{q}_2} \cdot \vec{R}_{q_2\bar{q}_2})
    \nonumber    \\
& = & \frac{3\sqrt{3}(m_{q_1}+m_{\bar{q}_2})^3(m_{q_2}+m_{\bar{q}_4})^3
(m_{q_3}+m_{\bar{q}_1})^3}
{\sqrt {V^5}(m_{q_1}m_{q_2}m_{q_3}+m_{\bar{q}_1}m_{\bar{q}_2}m_{\bar{q}_4})^3}
\int d\vec{r}_{q_1\bar{q}_1} d\vec{r}_{q_2\bar{q}_2} d\vec{R}_{\rm total} 
d\vec{\rho}_Y d\vec{\lambda}_Y
    \nonumber    \\
& & \psi_{q_1\bar{q}_2}^+ (\vec{r}_{q_1\bar{q}_2})
\psi_{q_2\bar{q}_4}^+ (\vec{r}_{q_2\bar{q}_4})
\psi_{q_3\bar{q}_1}^+ (\vec{r}_{q_3\bar{q}_1})
V_{{\rm D}_5}
\psi_{q_1\bar{q}_1}(\vec{r}_{q_1\bar{q}_1})
\psi_{q_2\bar{q}_2}(\vec{r}_{q_2\bar{q}_2})
    \nonumber    \\
& & \exp ( -i\vec{P}_{\rm f} \cdot \vec{R}_{\rm total}
-i\vec{p}_{\rho_Y} \cdot \vec{\rho}_Y
-i\vec{p}_{\lambda_Y} \cdot \vec{\lambda}_Y
+i\vec{P}_{\rm i} \cdot \vec{R}_{\rm total}
+i\vec{p}_{q_1\bar{q}_1,q_2\bar{q}_2}\cdot\vec{r}_{q_1\bar{q}_1,q_2\bar{q}_2})
    \nonumber    \\
& = & (2\pi)^3 \delta^3 (\vec{P}_{\rm i} - \vec{P}_{\rm f}) 
\frac {{\cal M}_{{\rm D}_5} }
{\sqrt {V^5}\sqrt{2E_A2E_B2E_{C_1}2E_{C_2}2E_{C_3}}},
\end{eqnarray}
where ${\cal M}_{{\rm D}_5}$ is the transition amplitude given by
\begin{eqnarray}
{\cal M}_{{\rm D}_5}
& = & \sqrt {2E_A2E_B2E_{C_1}2E_{C_2}2E_{C_3}}
\frac{3\sqrt{3}(m_{q_1}+m_{\bar{q}_2})^3(m_{q_2}+m_{\bar{q}_4})^3
(m_{q_3}+m_{\bar{q}_1})^3}
{(m_{q_1}m_{q_2}m_{q_3}+m_{\bar{q}_1}m_{\bar{q}_2}m_{\bar{q}_4})^3}
         \nonumber    \\
& & \int d\vec{r}_{q_1\bar{q}_1}d\vec{r}_{q_2\bar{q}_2}
d\vec{\rho}_Y d\vec{\lambda}_Y
\psi_{q_1\bar{q}_2}^+ (\vec{r}_{q_1\bar{q}_2})
\psi_{q_2\bar{q}_4}^+ (\vec{r}_{q_2\bar{q}_4})
\psi_{q_3\bar{q}_1}^+ (\vec{r}_{q_3\bar{q}_1}) V_{{\rm D}_5}
         \nonumber   \\
& & \psi_{q_1\bar{q}_1}(\vec{r}_{q_1\bar{q}_1})
\psi_{q_2\bar{q}_2}(\vec{r}_{q_2\bar{q}_2}) 
\exp  (-i\vec{p}_{\rho_Y} \cdot \vec{\rho}_Y
-i\vec{p}_{\lambda_Y} \cdot \vec{\lambda}_Y
+i\vec{p}_{q_1\bar{q}_1,q_2\bar{q}_2}\cdot\vec{r}_{q_1\bar{q}_1,q_2\bar{q}_2}).
         \nonumber   \\
\end{eqnarray}
In the above two equations the variable $\vec{\rho}_Y$ is defined from the 
position vectors of the two mesons with equal masses, and the variable 
$\vec{\lambda}_Y$ from the other meson. In case that
mesons $C_1(q_1\bar{q}_2)$ and $C_2(q_2\bar{q}_4)$ have the same mass,
we define
\begin{equation}
\vec{\rho}_Y=\frac{1}{\sqrt 2} (\vec{R}_{q_1\bar{q}_2}-\vec{R}_{q_2\bar{q}_4}),
\end{equation}
\begin{equation}
\vec{\lambda}_Y=\frac{1}{\sqrt 6} (\vec{R}_{q_1\bar{q}_2}
+\vec{R}_{q_2\bar{q}_4}-2\vec{R}_{q_3\bar{q}_1}).
\end{equation}
Let $\vec{p}_{\rho_Y}$ ($\vec{p}_{\lambda_Y}$) be $m_\rho$ ($m_\lambda$) times
the derivative of $\vec{\rho}_Y$ ($\vec{\lambda}_Y$) with respect to time.
In Eq. (29) we have used the two equalities,
\begin{equation}
d\vec{R}_{q_1\bar{q}_2}d\vec{R}_{q_2\bar{q}_4}d\vec{R}_{q_3\bar{q}_1}
=3\sqrt{3}d\vec{R}_{\rm total}d\vec{\rho}_Yd\vec{\lambda}_Y,
\end{equation}
\begin{equation}
\vec{P}_{q_1\bar{q}_2} \cdot \vec{R}_{q_1\bar{q}_2}
+\vec{P}_{q_2\bar{q}_4} \cdot \vec{R}_{q_2\bar{q}_4}
+\vec{P}_{q_3\bar{q}_1} \cdot \vec{R}_{q_3\bar{q}_1}
=\vec{P}_{\rm f} \cdot \vec{R}_{\rm total}
+\vec{p}_{\rho_Y} \cdot \vec{\rho}_Y+\vec{p}_{\lambda_Y} \cdot \vec{\lambda}_Y.
\end{equation}

For diagram ${\rm D}_6$ we have
\begin{eqnarray}
<C_1, C_2, C_3 \mid V_{{\rm D}_6}\mid A,B>
& = & <q_1\bar{q}_2, q_2\bar{q}_4, q_3\bar{q}_1 \mid 
V_{{\rm D}_6} \mid q_1\bar{q}_1, q_2\bar{q}_2>
    \nonumber    \\
& = & (2\pi)^3 \delta^3 (\vec{P}_{\rm i}-\vec{P}_{\rm f}) 
\frac {{\cal M}_{{\rm D}_6} }
{\sqrt {V^5}\sqrt{2E_A2E_B2E_{C_1}2E_{C_2}2E_{C_3}}},
\end{eqnarray}
where the transition amplitude ${\cal M}_{{\rm D}_6}$ is 
obtained from Eq. (30) by replacing $V_{{\rm D}_5}$ with 
$V_{{\rm D}_6}$. For diagram ${\rm D}_7$ we have
\begin{eqnarray}
<C_1, C_2, C_3 \mid V_{{\rm D}_7}\mid A,B>
& = & <q_1\bar{q}_2, q_2\bar{q}_4, q_3\bar{q}_1 \mid 
V_{{\rm D}_7} \mid q_1\bar{q}_1, q_2\bar{q}_2>
    \nonumber    \\
& = & (2\pi)^3 \delta^3 (\vec{P}_{\rm i}-\vec{P}_{\rm f}) 
\frac {{\cal M}_{{\rm D}_7} }
{\sqrt {V^5}\sqrt{2E_A2E_B2E_{C_1}2E_{C_2}2E_{C_3}}},
\end{eqnarray}
where the transition amplitude ${\cal M}_{{\rm D}_7}$ is 
obtained from Eq. (30) by replacing $V_{{\rm D}_5}$ with 
$V_{{\rm D}_7}$. For diagram ${\rm D}_8$ we have
\begin{eqnarray}
<C_1, C_2, C_3 \mid V_{{\rm D}_8}\mid A,B>
& = & <q_1\bar{q}_2, q_2\bar{q}_4, q_3\bar{q}_1 \mid 
V_{{\rm D}_8} \mid q_1\bar{q}_1, q_2\bar{q}_2>
    \nonumber    \\
& = & (2\pi)^3 \delta^3 (\vec{P}_{\rm i}-\vec{P}_{\rm f}) 
\frac {{\cal M}_{{\rm D}_8} }
{\sqrt {V^5}\sqrt{2E_A2E_B2E_{C_1}2E_{C_2}2E_{C_3}}},
\end{eqnarray}
where the transition amplitude ${\cal M}_{{\rm D}_8}$ is 
obtained from Eq. (30) by replacing $V_{{\rm D}_5}$ with 
$V_{{\rm D}_8}$.

Let $P_A$ ($P_B$) and $m_A$ ($m_B$) be the four-momentum and the
mass of meson $A$ ($B$), respectively, and we have the Mandelstam variable
$s=(P_A+P_B)^2$. Along the general lines provided in Ref. \cite{BD} on  
deriving the cross section from the transition amplitude, we get
the unpolarized cross section for $A+B \to C_1+C_2+C_3$,
\begin{eqnarray}
\sigma^{\rm unpol} & = &
\frac {(2\pi)^4}{4\sqrt {(P_A \cdot P_B)^2 - m_A^2m_B^2}}
\frac {1}{(2J_A+1)(2J_B+1)}
      \nonumber  \\
& & \int \frac {d^3P_{C_1}}{(2\pi)^3 2E_{C_1}} 
\frac {d^3P_{C_2}}{(2\pi)^3 2E_{C_2}} \frac {d^3P_{C_3}}{(2\pi)^3 2E_{C_3}}
\delta (E_{\rm f}-E_{\rm i}) \delta^3 (\vec{P}_{\rm f}-\vec{P}_{\rm i})
      \nonumber  \\
& & \sum\limits_{J_{Az}J_{Bz}J_{C_1z}J_{C_2z}J_{C_3z}}
\mid {\cal M}_{{\rm D}_1}+{\cal M}_{{\rm D}_2}
+{\cal M}_{{\rm D}_3}+{{\cal M}_{{\rm D}_4}}
+{\cal M}_{{\rm D}_5}+{\cal M}_{{\rm D}_6}
      \nonumber  \\
& & +{\cal M}_{{\rm D}_7}+{{\cal M}_{{\rm D}_8}}
\mid^2 ,
\end{eqnarray}
where $J_i$ ($i=A,B,C_1,C_2,C_3$) is the angular momentum of meson $i$ with
the magnetic projection quantum number $J_{iz}$. With the equality
$(P_A \cdot P_B)^2-m_A^2 m_B^2 =0.25 [s-(m_A+m_B)^2][s-(m_A-m_B)^2]$,
integration over $\vec{P}_{C_2}$ leads to
\begin{eqnarray}
\sigma^{\rm unpol} & = &
\frac {1}{16(2\pi)^5\sqrt {[s-(m_A+m_B)^2][s-(m_A-m_B)^2]}}
\frac {1}{(2J_A+1)(2J_B+1)}
      \nonumber  \\
& & \int \frac {d^3P_{C_1}d^3P_{C_3}}{E_{C_1}E_{C_2}E_{C_3}} 
\delta (E_{\rm f}-E_{\rm i}) 
\sum\limits_{J_{Az}J_{Bz}J_{C_1z}J_{C_2z}J_{C_3z}}
\mid {\cal M}_{{\rm D}_1}+{\cal M}_{{\rm D}_2}
      \nonumber  \\
& & +{\cal M}_{{\rm D}_3}+{{\cal M}_{{\rm D}_4}}
+{\cal M}_{{\rm D}_5}+{\cal M}_{{\rm D}_6}
+{\cal M}_{{\rm D}_7}+{{\cal M}_{{\rm D}_8}}
\mid^2 .
\end{eqnarray}
Integration over $\mid \vec{P}_{C_1} \mid$ yields
\begin{eqnarray}
\sigma^{\rm unpol} & = &
\frac {1}{16(2\pi)^5\sqrt {[s-(m_A+m_B)^2][s-(m_A-m_B)^2]}}
\frac {1}{(2J_A+1)(2J_B+1)}
      \nonumber  \\
& & \int \frac {d^3P_{C_3}}{E_{C_3}}d\Omega_{C_1} 
\frac{\mid \vec{P}_{C_1} \mid^2_0}
{\mid \mid \vec{P}_{C_1} \mid_0 E_{C_2}+(\mid \vec{P}_{C_1} \mid_0
-\mid \vec{P}_A+\vec{P}_B-\vec{P}_{C_3} \mid \cos \Theta )E_{C_1} \mid }
      \nonumber  \\
& & \sum\limits_{J_{Az}J_{Bz}J_{C_1z}J_{C_2z}J_{C_3z}}
\mid {\cal M}_{{\rm D}_1}+{\cal M}_{{\rm D}_2}
+{\cal M}_{{\rm D}_3}+{{\cal M}_{{\rm D}_4}}
+{\cal M}_{{\rm D}_5}+{\cal M}_{{\rm D}_6}
      \nonumber  \\
& & +{\cal M}_{{\rm D}_7}+{{\cal M}_{{\rm D}_8}}
\mid^2 ,
\end{eqnarray}
where $\Theta$ is the angle between $\vec{P}_{C_1}$ and 
$\vec{P}_A+\vec{P}_B-\vec{P}_{C_3}$, $d\Omega_{C_1}$ is the solid angle
centered about the direction of $\vec{P}_{C_1}$, and 
$\mid \vec{P}_{C_1} \mid_0$ is the absolute value of $\vec{P}_{C_1}$ that
satisfies $E_{\rm f}-E_{\rm i}=0$. The unpolarized cross section is a function
of $\sqrt s$, which is the total energy of the two initial mesons in the
center-of-mass frame.

\vspace{0.5cm}
\leftline{\bf III. TRANSITION POTENTIAL}
\vspace{0.5cm}

In Fig. 3 the left diagram denotes the process
$q^\prime (p_1) \to q^\prime (p_1^\prime) + q(p_3) + \bar{q}(-p_4)$,
and the right diagram $\bar{q}^\prime (-p_1) 
\to \bar{q}^\prime (-p_1^\prime) + q(p_3) + \bar{q}(-p_4)$.
In each diagram the wavy line represents the gluon which has  
four-momentum $k$, the color index $e$, and the space-time index $\tau$.
Each vertex involves the gauge coupling constant $g_{\rm s}$, 
the $SU(3)$ color generators $T^e$ $(e=1,\cdot \cdot \cdot,8)$, and
the Dirac matrices $\gamma^\tau$.
According to the Feynman rules in QCD \cite{Muta}, the amplitude 
for the left diagram in Fig. 3 is written as
\begin{equation}
{\cal M}_{{\rm c}q^\prime q\bar{q}}=\frac{{g_s}^2}{k^2}\bar{\psi}_{q'}
(\vec{p}^{~\prime}_1,s^{\prime}_{q'z})\gamma_\tau T^e\psi_{q'}
(\vec{p}_1,s_{q'z})
\bar{\psi}_q (\vec{p}_3,s_{qz}) \gamma^\tau T^e 
\psi_{\bar q}(\vec{p}_4,s_{\bar{q}z}),
\end{equation}
and the amplitude for the right diagram is 
\begin{equation}
{\cal M}_{{\rm c}\bar{q}^{\prime}q\bar{q}}=-\frac{{g_s}^2}{k^2}
{\bar{\psi}_{\bar{q}^{\prime}}}(\vec{p}_1,{s_{\bar{q}'z}})\gamma_{\tau} T^e
\psi_{\bar{q}'}(\vec{p}^{~\prime}_1,s^{\prime}_{\bar{q}'z})
\bar{\psi}_q(\vec{p}_3,s_{qz}) \gamma^{\tau} T^e
\psi_{\bar q}(\vec{p}_4,s_{\bar{q}z}),
\end{equation}
where repeated color and space-time
indices ($e$ and $\tau$) are summed. The quark spinors 
($\psi_{q'}(\vec{p}_1,s_{q'z})$, 
$\psi_{q'}(\vec{p}^{~\prime}_1,s^{\prime}_{q'z})$,
$\psi_q (\vec{p}_3,s_{qz})$) and the antiquark spinors
($\psi_{\bar q} (\vec{p}_4,s_{\bar{q}z})$, 
$\psi_{\bar{q}'}(\vec{p}_1,s_{\bar{q}'z})$,
$\psi_{\bar{q}'}(\vec{p}^{~\prime}_1,s^{\prime}_{\bar{q}'z})$) 
are given by \cite{SXW,BD}
\begin{equation}
\psi_q(\vec{p}_3,s_{qz})=\left(
\begin{array}{ccc} 
G_3(\vec {p}_3)\\{\frac{\vec\sigma\cdot\vec {p}_3}{2m_q}}G_3(\vec{p}_3)
\end{array}
\right)\chi_{s_{qz}},
\end{equation}
\begin{equation}
\psi_{\bar q}(\vec{p}_4,s_{\bar{q}z})=\left(
\begin{array}{ccc}
\frac{\vec\sigma\cdot\vec{p}_4}{2m_{\bar q}}G_4(\vec{p}_4)\\G_4(\vec{p}_4)
\end{array}
\right)\chi_{s_{\bar{q}z}},
\end{equation}
\begin{equation}
\psi_{q^{\prime}}(\vec{p}_1,s_{q'z})=\left(
\begin{array}{ccc}
G_1(\vec{p}_1)\\{\frac{\vec\sigma\cdot\vec{p}_1}{2m_{q'}}}G_1(\vec{p}_1)
\end{array}
\right)\chi_{s_{q'z}},
\end{equation}
\begin{equation}
\psi_{q'}(\vec{p}^{~\prime}_1,s^{\prime}_{q'z})=\left(
\begin{array}{ccc}
G^{\prime}_1(\vec{p}^{~\prime}_1)\\{\frac{\vec\sigma\cdot\vec{p}^{~\prime}_1}
{2m_{q'}}} G^{\prime}_1(\vec{p}^{~\prime}_1)
\end{array}
\right)\chi_{s^{\prime}_{q'z}},
\end{equation}
\begin{equation}
\psi_{\bar{q}'}(\vec{p}_1,s_{\bar{q}'z})=\left(
\begin{array}{ccc}
\frac{\vec\sigma\cdot\vec p_1}{2m_{\bar{q}'}}G_1(\vec {p}_1)\\G_1(\vec {p}_1)
\end{array}
\right)\chi_{s_{\bar{q}'z}},
\end{equation}
\begin{equation}
\psi_{\bar{q}'}(\vec{p}^{~\prime}_1,s^{\prime}_{\bar{q}'z})=\left(
\begin{array}{ccc}
\frac{\vec\sigma\cdot\vec{p}^{~\prime}_1}{2m_{\bar{q}'}}
G^{\prime}_1(\vec{p}^{~\prime}_1)\\
G^{\prime}_1(\vec {p}^{~\prime}_1)
\end{array}
\right)\chi_{s^{\prime}_{\bar{q}'z}},
\end{equation}
where $\vec \sigma$ are the Pauli matrices;
$\chi_{s_{qz}}$, $\chi_{s_{\bar{q}z}}$, $\chi_{s_{q'z}}$, 
$\chi_{s^{\prime}_{q'z}}$, $\chi_{s_{\bar{q}'z}}$, 
and $\chi_{s^{\prime}_{\bar{q}'z}}$
are the spin wave functions with the magnetic
projection quantum numbers, $s_{qz}$, $s_{\bar{q}z}$, $s_{q'z}$,
$s^{\prime}_{q'z}$, $s_{\bar{q}'z}$, and $s^{\prime}_{\bar{q}'z}$,
of the quark or antiquark spin, respectively. The quark and the antiquark 
created from the gluon have the same mass, i.e., $m_q=m_{\bar q}$.

Keeping terms to order of the inverse of the quark mass, we get
\begin{eqnarray}
{\cal M}_{{\rm c}q'q\bar{q}}&=&\frac{g_s^2}{k^2}\chi^+_{s^{\prime}_{q'z}}
\chi^+_{s_{qz}}T^eT^eG^{\prime}_1(\vec {p}^{~\prime}_1)G_3(\vec{p}_3)
\nonumber\\
& &
[\frac{\vec\sigma(34)\cdot\vec k}{2m_q}-\frac{\vec\sigma(1)\cdot\vec\sigma(34)
\vec\sigma(1)\cdot\vec {p}_1+\vec\sigma(1)\cdot\vec{p}^{~\prime}_1\vec\sigma(1)
\cdot\vec\sigma(34)}{2m_{q'}}]
\nonumber\\
& &
G_1(\vec {p}_1)G_4(\vec {p}_4)\chi_{s_{q'z}}\chi_{s_{\bar{q}z}} ,
\end{eqnarray}
\begin{eqnarray}
{\cal M}_{{\rm c}\bar{q}'q\bar{q}}&=&-\frac{g_s^2}{k^2}
\chi^+_{s_{\bar{q}^{\prime}z}}
\chi^+_{s_{qz}}T^eT^eG_1(\vec {p}_1)G_3(\vec {p}_3)
\nonumber\\
& &
[\frac{\vec\sigma(34)\cdot\vec k}{2m_q}
-\frac{\vec\sigma(1)\cdot\vec{p}_1\vec\sigma(1)\cdot\vec\sigma(34)
+\vec\sigma(1)\cdot\vec\sigma(34)\vec\sigma(1)\cdot\vec {p}^{~\prime}_1}
{2m_{\bar{q}'}}]
\nonumber\\
& &
G^{\prime}_1(\vec {p}^{~\prime}_1)G_4(\vec {p}_4)\chi_{s^{\prime}_{\bar{q}'z}}
\chi_{s_{\bar{q}z}} .
\end{eqnarray}
Using $T^eT^e=\frac{\vec{\lambda}(1)}{2}\cdot\frac{\vec{\lambda}(34)}{2}$ 
with $\vec \lambda$ being the Gell-Mann matrices, we
obtain the transition potential for $q^\prime (p_1)
\to {q}^\prime (p_1^\prime) + q(p_3) + \bar{q}(-p_4)$,
\begin{equation}
V_{{\rm c}q'q\bar{q}}(\vec{k}) =
\frac{\vec{\lambda}(1)}{2}\cdot\frac{\vec{\lambda}(34)}{2}
\frac{g_{\rm s}^2}{k^2}
\left(\frac{\vec{\sigma}(34)\cdot\vec{k}}{2m_q}-
\frac{\vec{\sigma}(1)\cdot\vec{\sigma}(34)\vec{\sigma}(1)\cdot\vec{p}_1+
\vec{\sigma}(1)\cdot\vec{p}_1^{~\prime} \vec{\sigma}(1)\cdot\vec{\sigma}(34)}
{2m_{q^\prime}}\right) ,
\end{equation}
and the transition potential for $\bar{q}^\prime (-p_1)
\to \bar{q}^\prime (-p_1^\prime)+ q(p_3) + \bar{q}(-p_4)$,
\begin{equation}
V_{{\rm c}\bar{q}'q\bar{q}}(\vec{k}) = -
\frac{\vec{\lambda}(1)}{2}\cdot\frac{\vec{\lambda}(34)}{2}
\frac{g_{\rm s}^2}{k^2}
\left(\frac{\vec{\sigma}(34)\cdot\vec{k}}{2m_q}-
\frac{\vec{\sigma}(1)\cdot\vec{p}_1 \vec{\sigma}(1)\cdot\vec{\sigma}(34)
+\vec{\sigma}(1)\cdot\vec{\sigma}(34)\vec{\sigma}(1)\cdot\vec{p}_1^{~\prime} }
{2m_{\bar{q}^\prime}}\right) .
\end{equation}
In Eqs. (51) and (52), $\vec{\lambda}(34)$ ($\vec{\sigma}(34)$)
mean that they have matrix elements 
between the color (spin) wave functions of the final quark
and the final antiquark.
In Eq. (51), $\vec{\lambda}(1)$ ($\vec{\sigma}(1)$) mean that they 
have matrix elements between the color (spin) wave functions of the final 
quark and the initial quark.
In Eq. (52), $\vec{\lambda}(1)$ ($\vec{\sigma}(1)$) mean that they 
have matrix elements between the color (spin) wave functions of the initial 
antiquark and the final antiquark. Applying Eqs. (51) and (52) to the eight
Feynman diagrams, we have
$V_{{\rm D}_1} \equiv V_{{\rm c}q_1q_3\bar{q}_4}$,
$V_{{\rm D}_2} \equiv V_{{\rm c}\bar{q}_1q_3\bar{q}_4}$,
$V_{{\rm D}_3} \equiv V_{{\rm c}q_2q_3\bar{q}_4}$,
$V_{{\rm D}_4} \equiv V_{{\rm c}\bar{q}_2q_3\bar{q}_4}$,
$V_{{\rm D}_5} \equiv V_{{\rm c}q_1q_3\bar{q}_4}$,
$V_{{\rm D}_6} \equiv V_{{\rm c}\bar{q}_1q_3\bar{q}_4}$,
$V_{{\rm D}_7} \equiv V_{{\rm c}q_2q_3\bar{q}_4}$, and
$V_{{\rm D}_8} \equiv V_{{\rm c}\bar{q}_2q_3\bar{q}_4}$.

\vspace{0.5cm}
\leftline{\bf IV. MATRIX ELEMENTS}
\vspace{0.5cm}

The transition amplitudes include color, spin, and flavor matrix elements. 
Denote the spin of meson $A$ ($B$, $C_1$, $C_2$, $C_3$) by $S_A$ 
($S_B$, $S_{C_1}$, $S_{C_2}$, $S_{C_3}$) and its magnetic projection quantum 
number 
by $S_{Az}$ ($S_{Bz}$, $S_{C_1z}$, $S_{C_2z}$, $S_{C_3z}$). Let
$\phi_{A\rm rel}$ ($\phi_{B\rm rel}$, $\phi_{C_1\rm rel}$, $\phi_{C_2\rm rel}$,
$\phi_{C_3\rm rel}$), $\phi_{A\rm color}$ ($\phi_{B\rm color}$, 
$\phi_{C_1\rm color}$, $\phi_{C_2\rm color}$, $\phi_{C_3\rm color}$), 
$\phi_{A\rm flavor}$ ($\phi_{B\rm flavor}$, 
$\phi_{C_1\rm flavor}$, $\phi_{C_2\rm flavor}$, $\phi_{C_3\rm flavor}$), 
and
$\chi_{S_A S_{Az}}$ ($\chi_{S_B S_{Bz}}$, $\chi_{S_{C_1}S_{C_1z}}$,
$\chi_{S_{C_2}S_{C_2z}}$, $\chi_{S_{C_3}S_{C_3z}}$)
be the quark-antiquark relative-motion wave function, the color wave function,
the flavor wave function, and the spin wave function of meson $A$ ($B$, $C_1$,
$C_2$, $C_3$), respectively. The wave function of mesons $A$ and $B$ is
\begin{equation}
\psi_{AB} =\phi_{A\rm rel} \phi_{B\rm rel} \phi_{A\rm color} \phi_{B\rm color}
\chi_{S_A S_{Az}} \chi_{S_B S_{Bz}} \varphi_{AB\rm flavor},
\end{equation}
and the wave function of mesons $C_1$, $C_2$, and $C_3$ is
\begin{equation}
\psi_{C_1C_2C_3} =\phi_{C_1\rm rel} \phi_{C_2\rm rel} \phi_{C_3\rm rel} 
\phi_{C_1\rm color} \phi_{C_2\rm color} \phi_{C_3\rm color} 
\chi_{S_{C_1}S_{C_1z}} \chi_{S_{C_2}S_{C_2z}} \chi_{S_{C_3}S_{C_3z}} 
\varphi_{C_1C_2C_3\rm flavor},
\end{equation}
where $\psi_{AB}=\psi_{q_1\bar{q}_1}\psi_{q_2\bar{q}_2}$ and
$\psi_{C_1C_2C_3}=\psi_{q_1\bar{q}_4}\psi_{q_2\bar{q}_1}\psi_{q_3\bar{q}_2}
=\psi_{q_1\bar{q}_2}\psi_{q_2\bar{q}_4}\psi_{q_3\bar{q}_1}$.
The flavor wave function $\varphi_{AB\rm flavor}$ of mesons $A$ 
and $B$ possesses the same isospin $I$ as the flavor wave function 
$\varphi_{C_1C_2C_3\rm flavor}$ of mesons $C_1$, $C_2$, and $C_3$. 

The color wave function of each meson is the color singlet. The color wave
function of mesons $A$ and $B$ is $\phi_{A\rm color} \phi_{B\rm color}$, and
the color wave function of mesons $C_1$, $C_2$, and $C_3$ is
$\phi_{C_1\rm color} \phi_{C_2\rm color} \phi_{C_3\rm color}$. The
color matrix element is
\begin{displaymath}
\phi_{C_1\rm color}^+ \phi_{C_2\rm color}^+ \phi_{C_3\rm color}^+
\frac{\vec{\lambda}}{2} \cdot \frac{\vec{\lambda}(34)}{2} 
\phi_{A\rm color} \phi_{B\rm color},
\end{displaymath}
where $\vec \lambda$ are the Gell-Mann matrices for the color generators of
quark $q_1$ in diagram $D_1$, antiquark $\bar{q}_1$ in diagram $D_2$, 
quark $q_2$ in diagram $D_3$, antiquark $\bar{q}_2$ in diagram $D_4$, 
quark $q_1$ in diagram $D_5$, antiquark $\bar{q}_1$ in diagram $D_6$, 
quark $q_2$ in diagram $D_7$, or antiquark $\bar{q}_2$ in diagram $D_8$. 
The color matrix element is $\frac {4}{9\sqrt 3}$,
$-\frac {4}{9\sqrt 3}$, $\frac {4}{9\sqrt 3}$, $-\frac {4}{9\sqrt 3}$,
$\frac {4}{9\sqrt 3}$, $-\frac {4}{9\sqrt 3}$, $\frac {4}{9\sqrt 3}$, and
$-\frac {4}{9\sqrt 3}$ for diagrams $D_1$, $D_2$, $D_3$, $D_4$, $D_5$, $D_6$, 
$D_7$, and $D_8$, respectively. 

The flavor wave functions, $\phi_{A\rm flavor}$ and $\phi_{B\rm flavor}$, are 
coupled to the flavor wave function $\varphi_{AB\rm flavor}$. The flavor
wave function of meson $C_1$ and the one of meson $C_2$ are coupled to the 
wave function $\varphi_{C_1C_2\rm flavor}$ of mesons $C_1$ and $C_2$ with the
total isospin $I^{\rm f}_{C_1C_2}$. Furthermore, 
$\varphi_{C_1C_2\rm flavor}$ and $\phi_{C_3\rm flavor}$ are coupled to
$\varphi_{C_1C_2C_3\rm flavor}$ with isospin $I$. Let
$P_{q_1 \to q_1+q_3+\bar{q}_4}$ ($P_{\bar{q}_1 \to \bar{q}_1+q_3+\bar{q}_4}$,
$P_{q_2 \to q_2+q_3+\bar{q}_4}$, $P_{\bar{q}_2 \to \bar{q}_2+q_3+\bar{q}_4}$)
denote the operator that implements 
$q_1 \to q_1+q_3+\bar{q}_4$ ($\bar{q}_1 \to \bar{q}_1+q_3+\bar{q}_4$,
$q_2 \to q_2+q_3+\bar{q}_4$, $\bar{q}_2 \to \bar{q}_2+q_3+\bar{q}_4$).
The flavor matrix elements corresponding to diagrams $D_1$, $D_2$, $D_3$, 
$D_4$, $D_5$, $D_6$, $D_7$, and $D_8$ are defined as
\begin{equation}
{\cal M}_{\rm D_1f}=\varphi^+_{C_1C_2C_3\rm flavor}
P_{q_1 \to q_1+q_3+\bar{q}_4}\varphi_{AB\rm flavor},
\end{equation}
\begin{equation}
{\cal M}_{\rm D_2f}=\varphi^+_{C_1C_2C_3\rm flavor}
P_{\bar{q}_1 \to \bar{q}_1+q_3+\bar{q}_4} \varphi_{AB\rm flavor},
\end{equation}
\begin{equation}
{\cal M}_{\rm D_3f}=\varphi^+_{C_1C_2C_3\rm flavor}
P_{q_2 \to q_2+q_3+\bar{q}_4}\varphi_{AB\rm flavor},
\end{equation}
\begin{equation}
{\cal M}_{\rm D_4f}=\varphi^+_{C_1C_2C_3\rm flavor}
P_{\bar{q}_2 \to \bar{q}_2+q_3+\bar{q}_4}\varphi_{AB\rm flavor},
\end{equation}
for $A(q_1\bar{q}_1)+B(q_2\bar{q}_2) \to C_1(q_1\bar{q}_4)+C_2(q_2\bar{q}_1)
+C_3(q_3\bar{q}_2)$ and
\begin{equation}
{\cal M}_{\rm D_5f}=\varphi^+_{C_1C_2C_3\rm flavor}
P_{q_1 \to q_1+q_3+\bar{q}_4}\varphi_{AB\rm flavor},
\end{equation}
\begin{equation}
{\cal M}_{\rm D_6f}=\varphi^+_{C_1C_2C_3\rm flavor}
P_{\bar{q}_1 \to \bar{q}_1+q_3+\bar{q}_4} \varphi_{AB\rm flavor},
\end{equation}
\begin{equation}
{\cal M}_{\rm D_7f}=\varphi^+_{C_1C_2C_3\rm flavor}
P_{q_2 \to q_2+q_3+\bar{q}_4}\varphi_{AB\rm flavor},
\end{equation}
\begin{equation}
{\cal M}_{\rm D_8f}=\varphi^+_{C_1C_2C_3\rm flavor}
P_{\bar{q}_2 \to \bar{q}_2+q_3+\bar{q}_4} \varphi_{AB\rm flavor},
\end{equation}
for $A(q_1\bar{q}_1)+B(q_2\bar{q}_2) \to C_1(q_1\bar{q}_2)+C_2(q_2\bar{q}_4)
+C_3(q_3\bar{q}_1)$. From the eight Feynman diagrams we have the relation,
\begin{equation}
{\cal M}_{\rm D_1f}={\cal M}_{\rm D_2f}
={\cal M}_{\rm D_3f}={\cal M}_{\rm D_4f},
\end{equation}
\begin{equation}
{\cal M}_{\rm D_5f}={\cal M}_{\rm D_6f}
={\cal M}_{\rm D_7f}={\cal M}_{\rm D_8f}.
\end{equation}
We list in Table 1 the flavor matrix elements for the following 2-to-3 
meson-meson reactions:
\begin{displaymath}
\pi \pi \to \pi K\bar{K},~\pi K \to \pi \pi K,~\pi K \to KK\bar{K},
~KK \to \pi KK,~K\bar{K} \to \pi K\bar{K},
\end{displaymath}
where $ K= \left( \begin{array}{c} K^+ \\ K^0 \end{array} \right) $ and
$\bar{K}= \left( \begin{array}{c} \bar{K}^0 \\ K^- \end{array} \right)$.

The initial mesons and the final mesons in the five reactions are pseudoscalar
mesons. The spin wave function of each pseudoscalar meson is the spin singlet
of the quark and the antiquark. The spin wave function of the two initial
mesons or the three final mesons is simply the product of the spin wave 
function of each meson
as seen in Eq. (53) or (54). Spin matrix elements are listed in Table 2. In
the table $\vec \sigma$ are the Pauli matrices for quark $q_1$ in diagrams
${\rm D}_1$ and ${\rm D}_5$, antiquark $\bar{q}_1$ in diagrams ${\rm D}_2$ 
and ${\rm D}_6$, quark $q_2$ in diagrams ${\rm D}_3$ and ${\rm D}_7$, 
or antiquark $\bar{q}_2$ in diagrams ${\rm D}_4$ and ${\rm D}_8$. 

The mesonic quark-antiquark relative-motion wave functions,
$\phi_{A\rm rel}$, $\phi_{B\rm rel}$, $\phi_{C_1\rm rel}$, 
$\phi_{C_2\rm rel}$, and $\phi_{C_3\rm rel}$, 
are the solutions of the Schr\"odinger equation with 
the potential \cite{JSX} between constituents $a$ and $b$:
\begin{eqnarray}
V_{ab}(\vec {r}) & = &
- \frac {\vec{\lambda}_a}{2} \cdot \frac {\vec{\lambda}_b}{2}
\frac {3}{4} D \left[ 1.3- \left( \frac {T}{T_{\rm c}} \right)^4 \right]
\tanh (Ar) + \frac {\vec{\lambda}_a}{2} \cdot \frac {\vec{\lambda}_b}{2}
\frac {6\pi}{25} \frac {v(\lambda r)}{r} \exp (-Er)
\nonumber  \\
& & -\frac {\vec{\lambda}_a}{2} \cdot \frac {\vec{\lambda}_b}{2}
\frac {16\pi^2}{25}\frac{d^3}{\pi^{3/2}}\exp(-d^2r^2) \frac {\vec {s}_a \cdot 
\vec {s} _b} {m_am_b}
+\frac {\vec{\lambda}_a}{2} \cdot \frac {\vec{\lambda}_b}{2}\frac {4\pi}{25}
\frac {1} {r}
\frac {d^2v(\lambda r)}{dr^2} \frac {\vec {s}_a \cdot \vec {s}_b}{m_am_b} ,
\end{eqnarray}
in which $D=0.7$ GeV, $E=0.6$ GeV, 
$\lambda=\sqrt{25/16\pi^2 \alpha'}$ with $\alpha'=1.04$ GeV$^{-2}$, and
$A=1.5[0.75+0.25 (T/{T_{\rm c}})^{10}]^6$ GeV, where $T$ is the temperature
and $T_{\rm c}$ is the critical temperature which equals 0.175 GeV \cite{KLP}.
The function $v$ is given by
Buchm\"uller and Tye in Ref. \cite{BT}, and the quantity $d$ is given by
\begin{eqnarray}
d^2=d_1^2\left[\frac{1}{2}+\frac{1}{2}\left(\frac{4m_a m_b}{(m_a+m_b)^2}
\right)^4\right]+d_2^2\left(\frac{2m_am_b}{m_a+m_b}\right)^2,
\end{eqnarray}
where $d_1=0.15$ GeV and $d_2=0.705$. The potential is a function 
of the distance $r$ between constituents $a$ and $b$, and contains the spins, 
$\vec {s}_a$ and $\vec {s}_b$, and the Gell-Mann matrices  
$\vec{\lambda}_a$ and $\vec{\lambda}_b$.
When the masses of the up quark, the down quark, the strange quark,
and the charm quark are 0.32 GeV, 0.32 GeV, 0.5 GeV, and 1.51 GeV, 
respectively, the meson masses obtained from the Schr\"odinger equation with
the potential at zero temperature reproduce
the experimental masses of $\pi$, $\rho$, $K$, $K^*$,
$J/\psi$, $\psi'$, $\chi_{c}$, $D$, $D^*$, $D_s$, and $D^*_s$ 
mesons~\cite{PDG}. Moreover, the experimental data of $S$-wave phase shifts
for elastic $\pi \pi$ scattering for $I=2$ in 
vacuum~\cite{CMSJS,LCFMP,Hoogland,DBGGT} are reproduced in the Born 
approximation.

\vspace{0.5cm}
\leftline{\bf V. NUMERICAL CROSS SECTIONS AND DISCUSSIONS}
\vspace{0.5cm}

We consider the following 2-to-3 meson-meson reactions:
\begin{displaymath}
\pi \pi \to \pi K\bar{K},~\pi K \to \pi \pi K,~\pi K \to KK\bar{K},
~KK \to \pi KK,~K\bar{K} \to \pi K\bar{K}.
\end{displaymath}
The reaction $\pi\bar{K} \to \pi\pi\bar{K}$ ($\pi\bar{K} \to K\bar{K}
\bar{K}$, $\bar{K}\bar{K} \to \pi \bar{K}\bar{K}$) has the same  
cross section as $\pi K \to \pi\pi K$ ($\pi K \to KK\bar{K}$, 
$KK \to \pi KK$). Cross sections for meson-meson reactions
depend on the flavor matrix 
elements. Based on the flavor matrix elements, cross sections for some
isospin channels of reactions can be obtained from the other isospin channels. 
Therefore, we calculate cross sections for the following eight channels:
\begin{displaymath}
I=2~I_{\pi K}^{\rm f}=\frac{3}{2}~\pi\pi \to \pi K\bar{K},
~I=1~I_{\pi K}^{\rm f}=\frac{3}{2}~\pi\pi \to \pi K\bar{K},
\end{displaymath}
\begin{displaymath}
I=\frac{3}{2}~I_{\pi K}^{\rm f}=\frac{3}{2}~\pi K \to \pi \pi K,
~I=\frac{3}{2}~I_{\pi K}^{\rm f}=\frac{1}{2}~\pi K \to \pi \pi K,
\end{displaymath}
\begin{displaymath}
I=\frac{3}{2}~I_{KK}^{\rm f}=1~\pi K \to KK\bar{K},
~I=1~I_{\pi K}^{\rm f}=\frac{3}{2}~KK \to \pi KK,
\end{displaymath}
\begin{displaymath}
I=1~I_{\pi K}^{\rm f}=\frac{1}{2}~KK \to \pi KK,
~I=1~I_{\pi \bar{K}}^{\rm f}=\frac{3}{2}~K\bar{K} \to \pi K\bar{K}.
\end{displaymath}
According to Eq.~(40)
we calculate the unpolarized cross section at the six temperatures
$T/T_{\rm c} =0$, $0.65$, $0.75$, $0.85$, $0.9$, and $0.95$. 
We plot the unpolarized cross sections for the eight channels of the reactions
in Figs. 4-11. These cross sections are functions of the temperature of 
hadronic matter and the Mandelstam variable $\sqrt s$. 

Every curve in Figs. 4-11 has a peak. Let $\sqrt{s_0}$ be the threshold energy.
Denote by $d_0$ the separation between the peak's location on
the $\sqrt s$-axis and the threshold energy. The
numerical cross sections shown in Figs. 4-11 are parametrized as
\begin{eqnarray}
\sigma^{\rm unpol}(\sqrt {s},T)
&=&a_1 \left( \frac {\sqrt {s} -\sqrt {s_0}} {b_1} \right)^{e_1}
\exp \left[ e_1 \left( 1-\frac {\sqrt {s} -\sqrt {s_0}} {b_1} \right) \right]
\nonumber \\
&&+ a_2 \left( \frac {\sqrt {s} -\sqrt {s_0}} {b_2} \right)^{e_2}
\exp \left[ e_2 \left( 1-\frac {\sqrt {s} -\sqrt {s_0}} {b_2} \right) \right].
\end{eqnarray}
The values of the parameters $a_1$, $b_1$, $e_1$, $a_2$,
$b_2$, and $e_2$ are listed in Tables 3 and 4, where $\sqrt{s_{\rm z}}$ is
the square root of the Mandelstam variable at which the cross section is
1/100 of the peak cross section. 

Since the sum of the masses of the final mesons is larger than the sum of the
masses of the initial mesons, the reactions are endothermic. When $\sqrt s$
increases from the threshold energy, which is the
sum of the masses of the final mesons, the cross section of every reaction
shown in Figs. 4-11 increases from zero to a maximum and then decreases.
The change of the peak cross section with temperature is obvious,
and the peak cross section at $T/T_{\rm c}=0.95$ is smallest among the peak
cross sections at the six temperatures.

As the temperature increases, values of the central spin-independent
potential [the first term and the second term of the right-hand side in 
Eq. (65)] at large distances become smaller and smaller (confinement
becomes weaker and weaker), and the Schr\"odinger equation produces increasing
meson radii. The weakening confinement with increasing temperature makes 
combining final quarks and antiquarks into mesons more difficult, and thus
reduces the cross sections. In contrast to decreasing peak cross sections 
caused by
weakening confinement, increasing peak cross sections are caused by
increasing radii of the initial mesons. When the decrease is faster than
the increase, the peak cross section goes down as the
temperature changes from a value (for example, 0.65$T_{\rm c}$ in Fig. 8
that show cross sections for $\pi K \to KK\bar K$ for $I=3/2$ and 
$I_{KK}^f=1$) to 0.95$T_{\rm c}$.

With increasing temperature, the meson radii increase. This corresponds to 
increasing wave functions at small quark-antiquark relative momentum. The
relative momentum depends linearly on the three-dimensional momentum $\vec P$
of an initial meson in the center-of-mass frame of the two initial mesons,
\begin{equation}
\vec{P}^2=\frac{1}{4}[s-(m_A+m_B)^2][1-\frac{(m_A-m_B)^2}{s}] .
\end{equation} 
The small relative momentum
may be given by small values of $\mid \vec{P} \mid$ and, furthermore,
of $\sqrt{s}$. A consequence is that $d_0$ listed in Tables 3 and 4
decreases or stays unchanged with
increasing temperature. The peak cross section occurs at 
$\sqrt{s}=m_{C_1}+m_{C_2}+m_{C_3}+d_0$. With increasing temperature, the
decrease of the pion and kaon masses \cite{LX} in addition to $d_0$
lead to the decrease of $\sqrt{s}=m_{C_1}+m_{C_2}+m_{C_3}+d_0$.

Cross sections for $\pi K \to \pi\pi K$ for $I=3/2$ were measured in Ref.
\cite{JMTVH}, but systematic and statistical
uncertainties were not given. The experimental cross
section is 0.04 mb at $\sqrt{s}=0.95$ GeV and 0.16 mb at $\sqrt{s}=1.15$ GeV.
The two data are individually near the values 0.014 mb and 0.2 mb of the 
present work at zero temperature.

The 2-to-3 meson-meson scattering is caused by 
a gluon created from a quark or an antiquark and the gluon creates a
quark-antiquark pair. If the quark-antiquark pair is $u\bar u$ or $d\bar d$,
we have the reaction $\pi K \to \pi\pi K$. If the quark-antiquark pair is 
$s\bar s$, we have the reaction $\pi K \to KK\bar K$. Since the up-quark
and down-quark masses are smaller than the strange-quark mass, it is more 
likely to create a $u\bar u$ or $d\bar d$ pair than a $s\bar s$ pair.
Therefore, the peak cross sections of $\pi K \to \pi\pi K$ for $I=3/2$ at a 
given temperature in Figs. 6 and 7 are larger than the one of 
$\pi K \to KK\bar K$ for $I=3/2$ and $I^{\rm f}_{KK}=1$ in Fig. 8.

Some 2-to-3 meson-meson reactions in the present work and some 2-to-2
meson-meson reactions in Ref. \cite{SXW} have the same initial
mesons. We can thus compare the cross sections obtained in the
present work and those provided in Ref. \cite{SXW}. At a given temperature
the peak cross section of
$\pi \pi \to \pi K \bar K$ for $I=1$ and $I^{\rm f}_{\pi K}=3/2$ in Fig. 5
is smaller than the one of $\pi \pi \to K \bar{K}^\ast$ for $I=1$ in Ref.
\cite{SXW}. Cross sections for $\pi K \to \pi\pi K$ for $I=3/2$ are shown 
in Figs. 6 and 7. According to the flavor matrix elements in Table 1, the 
cross section for $\pi K \to \pi\pi K$ for $I=1/2$ and $I^{\rm f}_{\pi K}=3/2$
is 1.6 times the one for $\pi K \to \pi\pi K$ for $I=3/2$ and 
$I^{\rm f}_{\pi K}=3/2$, and the cross section for $\pi K \to \pi\pi K$ for
$I=1/2$ and $I^{\rm f}_{\pi K}=1/2$ is 0.25 times the one for
$\pi K \to \pi\pi K$ for $I=3/2$ and $I^{\rm f}_{\pi K}=1/2$. The peak cross
section of $\pi K \to \pi\pi K$ for $I=1/2$ and $I^{\rm f}_{\pi K}=3/2$
is smaller than the one of $\pi K \to \pi K^*$ for $I=1/2$ at 
$T/T_{\rm c}=0$, $0.65$, and 0.75, but larger at $T/T_{\rm c}=0.85$, $0.9$,
and 0.95. The peak cross section of $\pi K \to \pi\pi K$ for $I=1/2$ and 
$I^{\rm f}_{\pi K}=1/2$ is smaller than the one of $\pi K \to \pi K^*$ for 
$I=1/2$. The peak cross section of $K\bar{K} \to \pi K \bar{K}$ for $I=1$
and $I^{\rm f}_{\pi \bar K}=3/2$ in Fig. 11 is smaller than the one of 
$K\bar{K} \to 
K\bar{K}^*$ for $I=1$ at $T/T_{\rm c}=0$, 0.65, 0.75, and 0.95, but larger at
$T/T_{\rm c}=0.85$ and 0.9. Therefore, 2-to-3 meson-meson scattering may be
as important as inelastic 2-to-2 meson-meson scattering.

The $\tau$ decay has been used to study asymptotic freedom of QCD 
\cite{BCK,DDHMZ}. The $\tau$ lepton decays into $\nu_\tau$ and
$W$ which splits into a quark and an antiquark. If the quark or the antiquark
emits a virtual gluon which subsequently splits into a quark-antiquark pair,
decay modes like $\tau^- \to \pi^- \bar{K}^0 \nu_\tau$ and 
$\tau^- \to K^- K^0 \nu_\tau$ are observed. If the quark and/or the antiquark
creates two virtual gluons of which each subsequently splits into a 
quark-antiquark pair, decay modes like $\tau^- \to \pi^- \bar{K}^0 \pi^0 
\nu_\tau$  and $\tau^- \to K^- K^0 \pi^0 \nu_\tau$ are observed. That
the virtual gluon splits into a quark-antiquark pair also takes place in 2-to-3
meson-meson scattering in the present work, and perturbative QCD is applied to
the process.

In perturbative QCD physical observables are usually given by a power series
in $\alpha_{\rm s}$, which is $g_s^2/4\pi$. 
If the coupling constant $\alpha_{\rm s}$ is smaller than 1, the perturbative
expansion converges. In the present work the coupling constant is 0.75 from 
Ref. \cite{BT}, and is used in the Feynman diagrams shown in Fig. 3.

When we add the gluon propagator, the gluon loop, the quark loop, and the ghost
loop
to the eight diagrams in Figs. 1 and 2, this generates 312 Feynman diagrams at 
order 
$\alpha_{\rm s}^2$. The 312 diagrams and the 8 diagrams in Figs. 1 and 2 do
not contain quark-antiquark annihilation. From the annihilation of an initial
quark and an initial antiquark as well as the creation of a quark-antiquark 
pair,
we get 14 Feynman diagrams at order $\alpha_{\rm s}^2$. In total, we have 326
diagrams. The calculation of such a large number of diagrams is formidable.

The transition potentials given in Eqs. (51) and (52) consist of terms with the
inverse of the quark mass. In obtaining the transition potentials the terms 
with
the inverse of the quark mass cubed are neglected, because they are suppressed
by the inverse of the quark mass squared in comparison to the terms in Eqs. 
(51) and (52).

Nonperturbative effects exist in 2-to-3 meson-meson scattering, and
are encoded in the mesonic quark-antiquark wave functions 
as done in Refs. \cite{KQS,GKS}. The quark-antiquark pair from the virtual 
gluon combines
with spectator quarks and antiquarks from the initial mesons to form three 
final mesons. The combination involves multi-gluon exchange between the quark
and the antiquark, and confinement sets in. While the final
mesons are formed, the wave functions are determined.

If two or more mesons are produced in a reaction or a decay, they interact with
each other before being detected. The role of final state interactions has been
studied in chiral perturbation theory. While two mesons are produced in a 
photon-photon reaction, final state interactions come from meson loops
and meson resonances between the two photons and the final mesons 
\cite{OO3,MNW}. In reproducing experimental data of $\gamma \gamma$ cross 
sections, final state interactions are essential. However, in the reaction
$\pi N \to \pi \pi N$ final state interactions cause a correction less than
20 \% when the total center-of-mass energy of initial $\pi N$ is 
smaller than 2 GeV \cite{AB}. In the hadronic decays $\eta \to 3\pi$, 
$\eta^\prime \to 3\pi$, and $\eta^\prime \to \eta \pi \pi$, final state 
interactions due to loop corrections and resonances lead to excellent agreement
of theoretical decay widths with experimental data, $\pi \pi$ rescattering
is shown to be important, but $S$-wave $\pi \eta$ rescattering effects are 
small \cite{BN,GP}. For decay modes like 
$D\to \pi \pi \pi$,
$D\to \bar{K} \pi \pi$, $D\to KK \bar K$, $J/\psi \to \pi \pi \pi$, 
$J/\psi \to \phi \pi \pi$, $J/\psi \to \phi K \bar K$, 
$\bar{B}^0 \to \pi \pi \pi$, $\bar{B}^0 \to J/\psi \pi \pi$,
$\bar{B}^0 \to J/\psi \pi \eta$, and $\bar{B}^0 \to J/\psi K\bar K$,
the decay amplitude is assumed to be linearly dependent on amplitudes of 
rescattering diagrams of two final mesons since the weak interaction is 
involved
\cite{Oller,Nakamura,AMRR,MP,MO,RPOC,GMS,GM,ADHKM}. Experimental data on these
decays may be accounted for.

In the meson-meson collisions that produce three mesons, final state 
interactions due to resonances and loop corrections exist. For instance,
the $K^*$ resonance contributes to $\pi \pi \to \pi K \bar K$ through
$\pi \pi \to K^* \bar{K}$ and $K^* \to \pi K$; $\pi K \to \pi \pi K$ may
happen through $\pi K \to \pi \rho K$ and $\rho K \to \pi K$.
If the final state interactions are taken into account, more accurate 
cross sections are expected, but we do not include the final state 
interactions in the present work. When the two initial mesons approach each
other, they undergo elastic scattering. The initial state interaction may
influence the production of the three final mesons, but we do not include the 
initial state interaction in the present work.

\vspace{0.5cm}
\leftline{\bf VI. SUMMARY }
\vspace{0.5cm}

We have proposed a model to study 2-to-3 meson-meson scattering.
A gluon is created from a quark or an antiquark constituent, and 
subsequently the gluon
splits into a quark and an antiquark. This process causes a meson-meson
collision to produce three mesons. The transition potential
for the process has been derived from the Feynman rules in perturbative QCD.
Eight Feynman diagrams at tree level are
involved in the 2-to-3 meson-meson scattering. 
From the $S$-matrix element we have derived the transition amplitudes
corresponding to the eight Feynman diagrams, and from the eight transition 
amplitudes we have derived the unpolarized cross section. The 2-to-3 reactions
among pions and kaons include 
$\pi \pi \to \pi K\bar{K}$, $\pi K \to \pi \pi K$, $\pi K \to KK\bar{K}$,
$KK \to \pi KK$, and $K\bar{K} \to \pi K\bar{K}$.
We have calculated color, spin, and flavor matrix elements for these 
reactions. From the cross-section formulas we have
obtained numerical unpolarized cross sections for eight isospin channels of the
reactions, and the numerical cross section results are parametrized. At zero
temperature our cross sections for $\pi K \to \pi\pi K$ for $I=3/2$ are near 
the experimental data. The unpolarized
cross sections depend on temperature. The cross section for any isospin 
channel at a given temperature has a maximum, and the peak cross section of 
any reaction decreases as the temperature approaches the critical
temperature. By comparison with inelastic
2-to-2 meson-meson scattering, we find that 2-to-3 meson-meson scattering
may be as important as inelastic 2-to-2 meson-meson scattering.

\vspace{0.5cm}
\leftline{\bf ACKNOWLEDGEMENTS}
\vspace{0.5cm}

This work was supported by the National Natural Science Foundation of China
under Grant No. 11175111.

\newpage
\begin{figure}[htbp]
  \centering
    \includegraphics[scale=0.82]{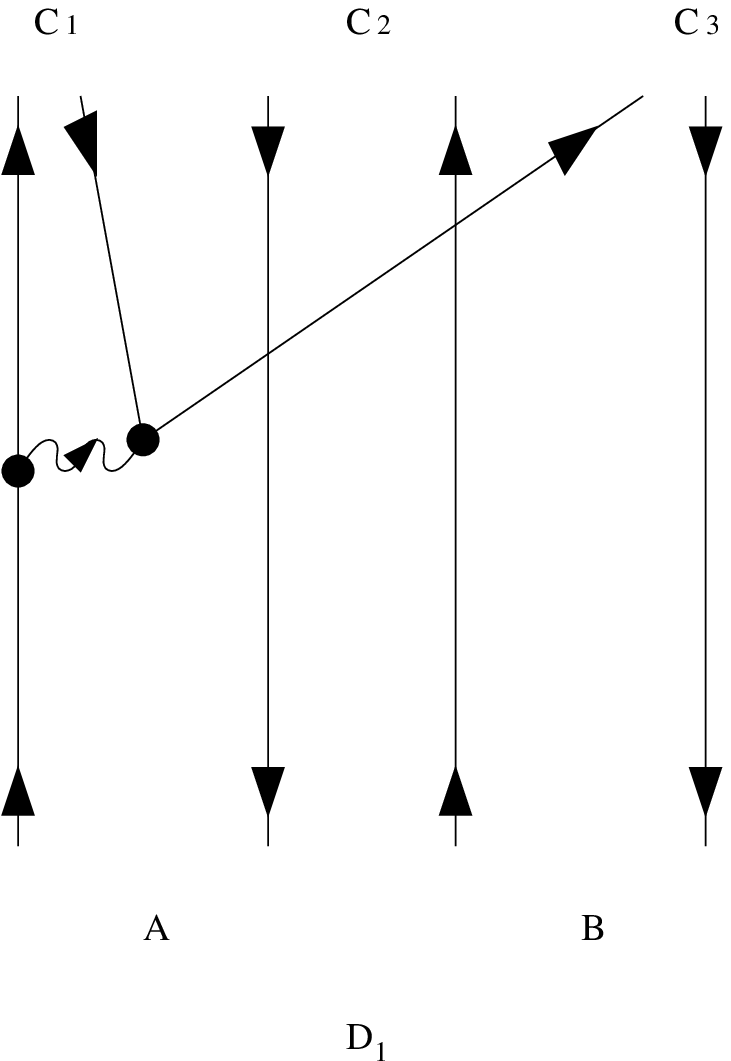}
      \hspace{3cm}
    \includegraphics[scale=0.82]{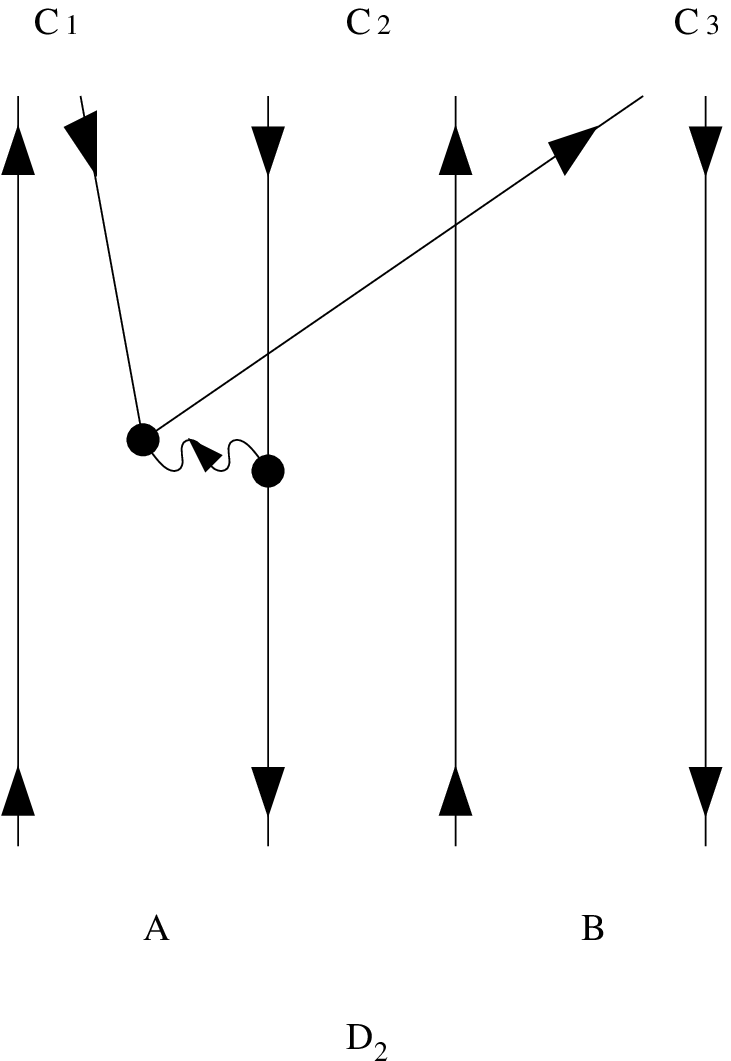}
      \vskip 72pt
    \includegraphics[scale=0.82]{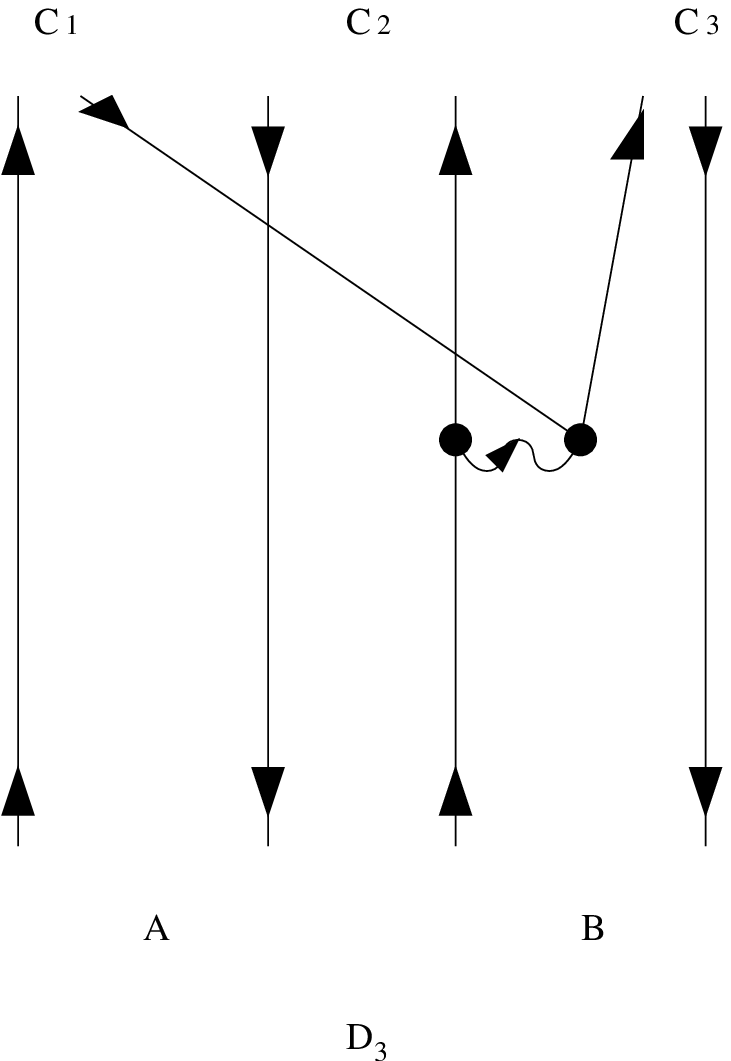}
      \hspace{3cm}
    \includegraphics[scale=0.82]{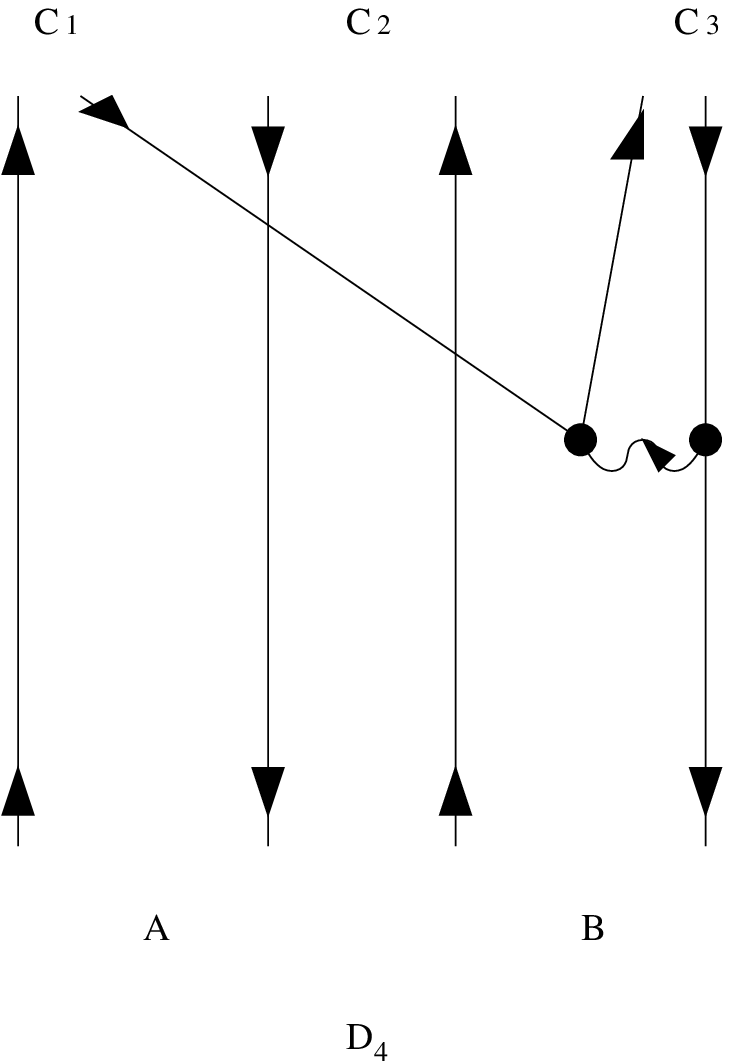}
\caption{Reaction $A(q_1\bar{q}_1)+B(q_2\bar{q}_2) \to 
C_1(q_1\bar{q}_4)+C_2(q_2\bar{q}_1)+C_3(q_3\bar{q}_2)$. 
Solid lines with up (down) arrows represent
quarks (antiquarks). Wavy lines represent gluons.}
\label{fig1}
\end{figure}

\newpage
\begin{figure}[htbp]
  \centering
    \includegraphics[scale=0.74]{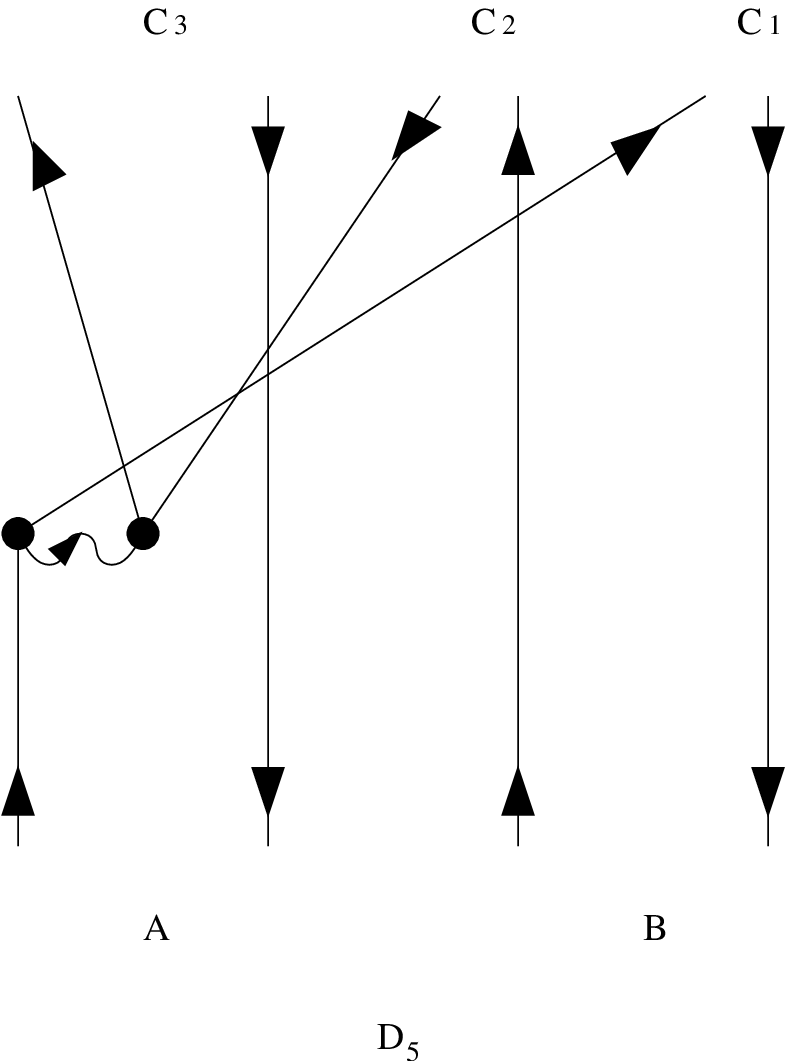}
      \hspace{3cm}
    \includegraphics[scale=0.74]{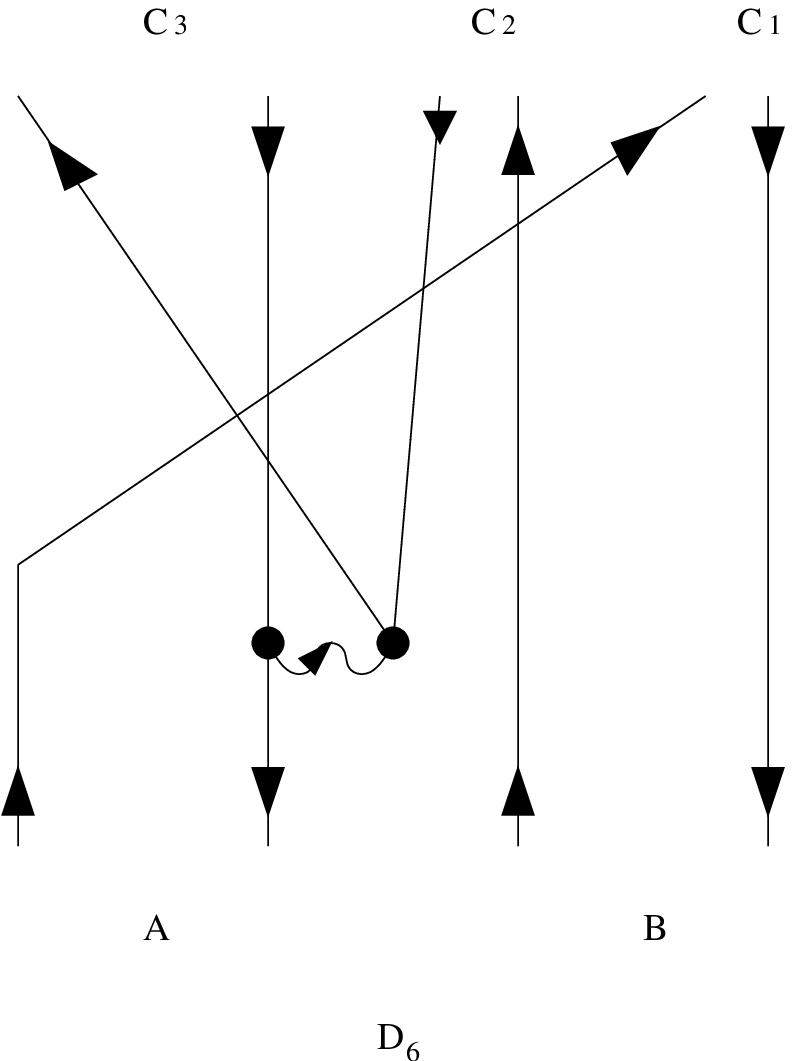}
      \vskip 72pt
    \includegraphics[scale=0.74]{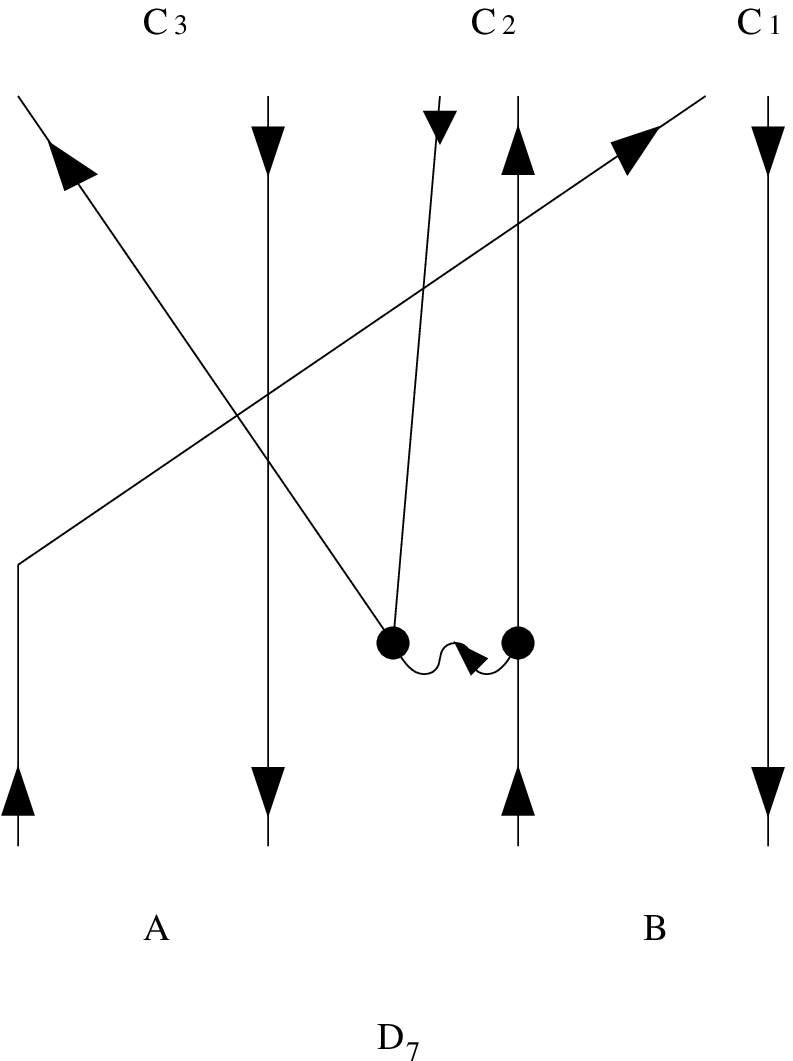}
      \hspace{3cm}
    \includegraphics[scale=0.74]{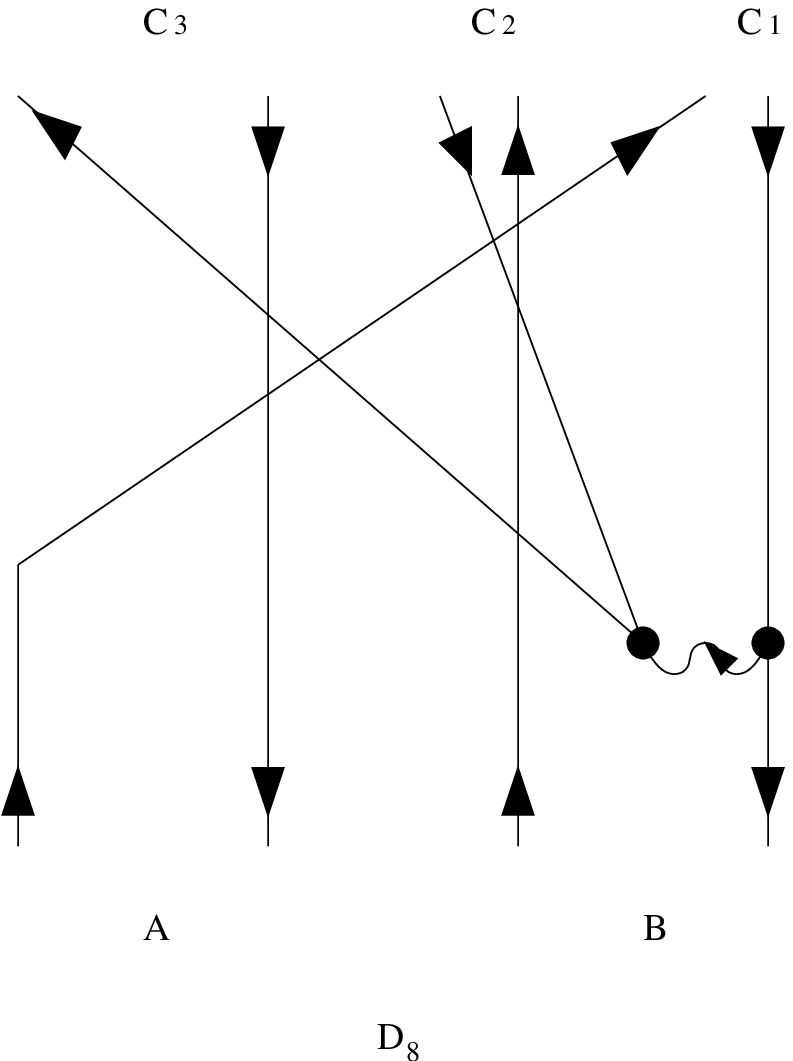}
\caption{Reaction $A(q_1\bar{q}_1)+B(q_2\bar{q}_2) \to 
C_1(q_1\bar{q}_2)+C_2(q_2\bar{q}_4)+C_3(q_3\bar{q}_1)$. 
Solid lines with up (down) arrows represent
quarks (antiquarks). Wavy lines represent gluons.}
\label{fig2}
\end{figure}

\newpage
\begin{figure}[htbp]
  \centering
    \includegraphics[scale=0.8]{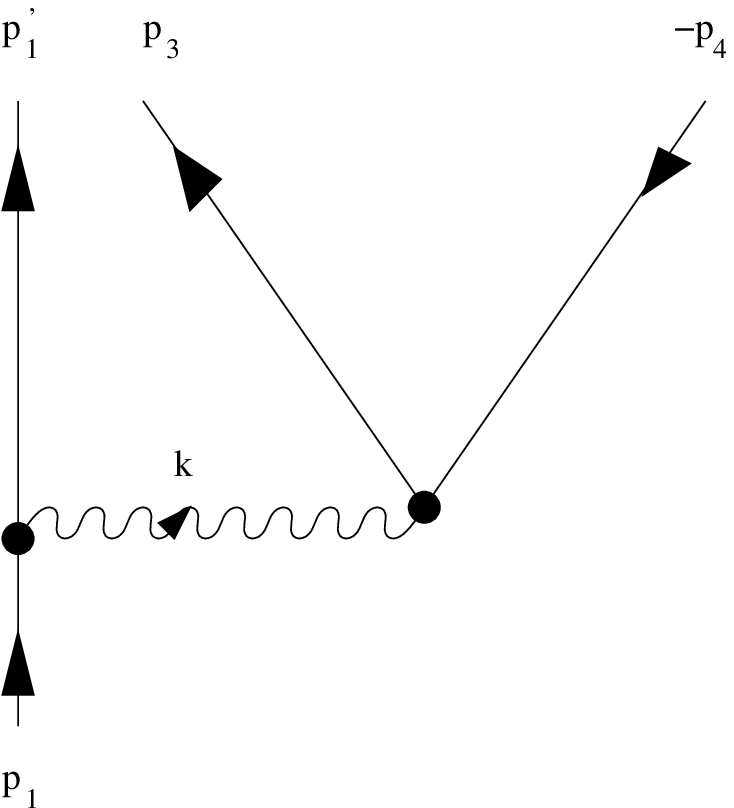}
      \hspace{2cm}
    \includegraphics[scale=0.8]{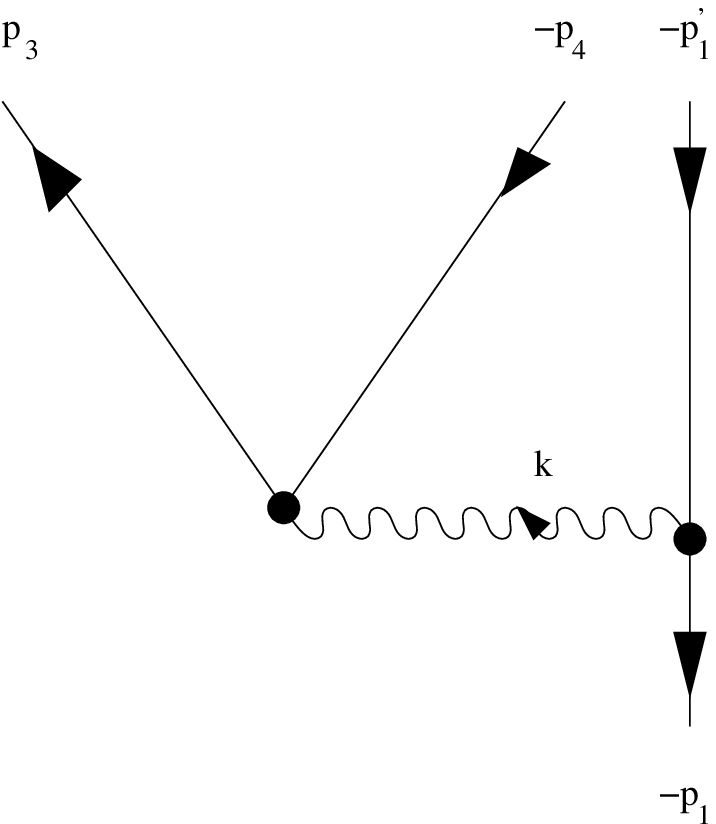}
\caption{Left diagram with $q'(p_1) \to q'(p_1^\prime)+q(p_3)+\bar{q}(-p_4)$ 
and right diagram with $\bar{q}'(-p_1) \to \bar{q}'(-p_1^\prime)+q(p_3)
+\bar {q}(-p_4)$.}
\label{fig3}
\end{figure}

\newpage
\begin{figure}[htbp]
  \centering
    \includegraphics[scale=0.55]{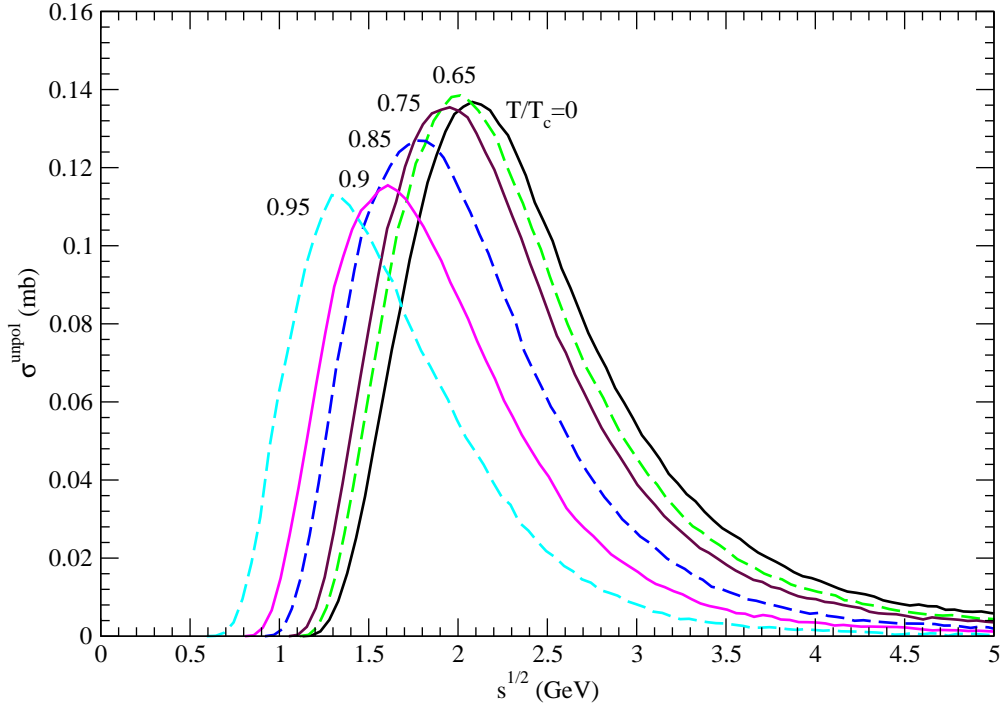}
\caption{Cross sections for $\pi \pi \to \pi K \bar K$ for $I=2$ and
$I_{\pi K}^f=3/2$ at various temperatures (in units of the critical 
temperature).}
\label{fig4}
\end{figure}

\newpage
\begin{figure}[htbp]
  \centering
    \includegraphics[scale=0.6]{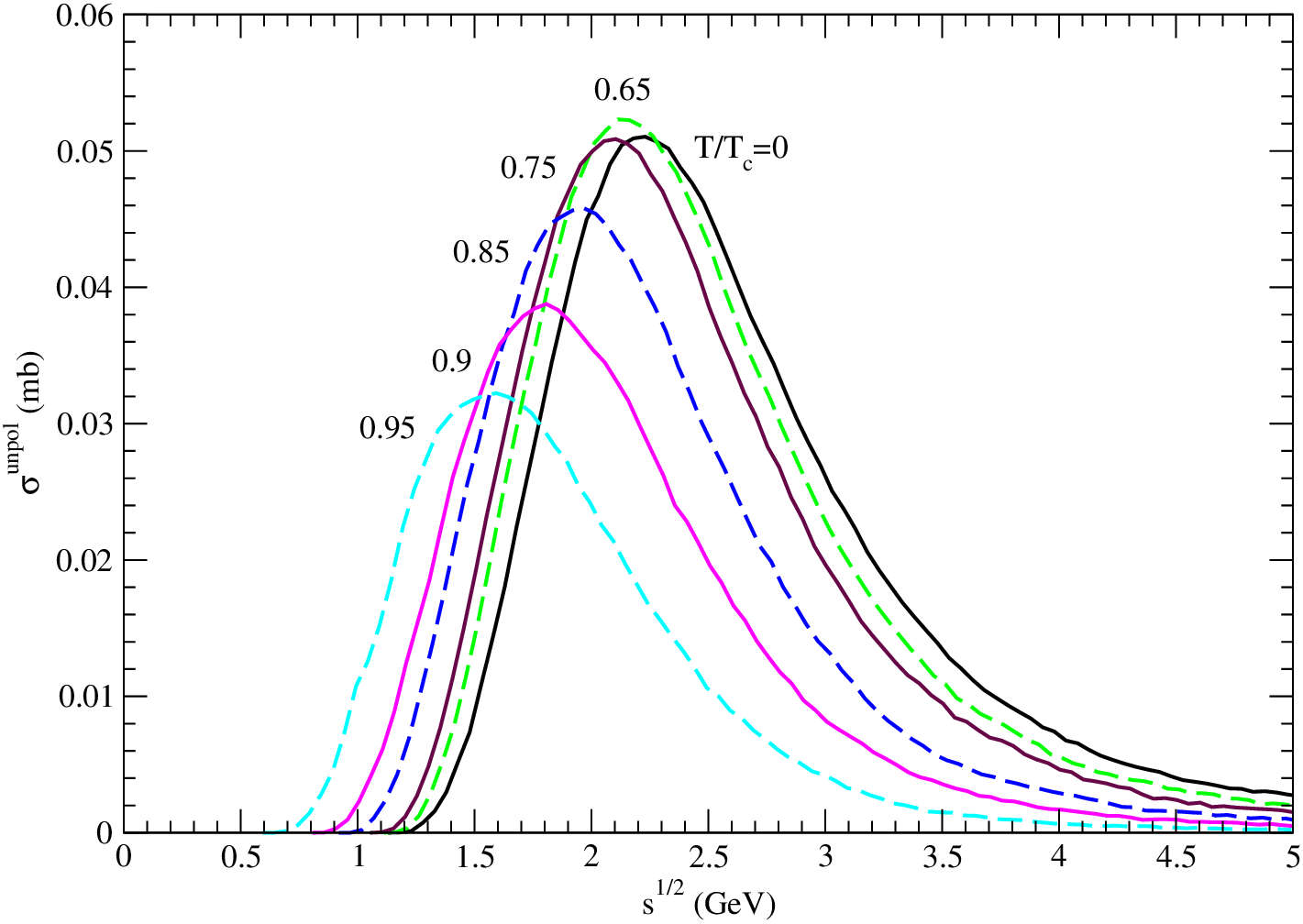}
\caption{Cross sections for $\pi \pi \to \pi K \bar K$ for $I=1$ and
$I_{\pi K}^f=3/2$ 
at various temperatures.}
\label{fig5}
\end{figure}

\newpage
\begin{figure}[htbp]
  \centering
    \includegraphics[scale=0.6]{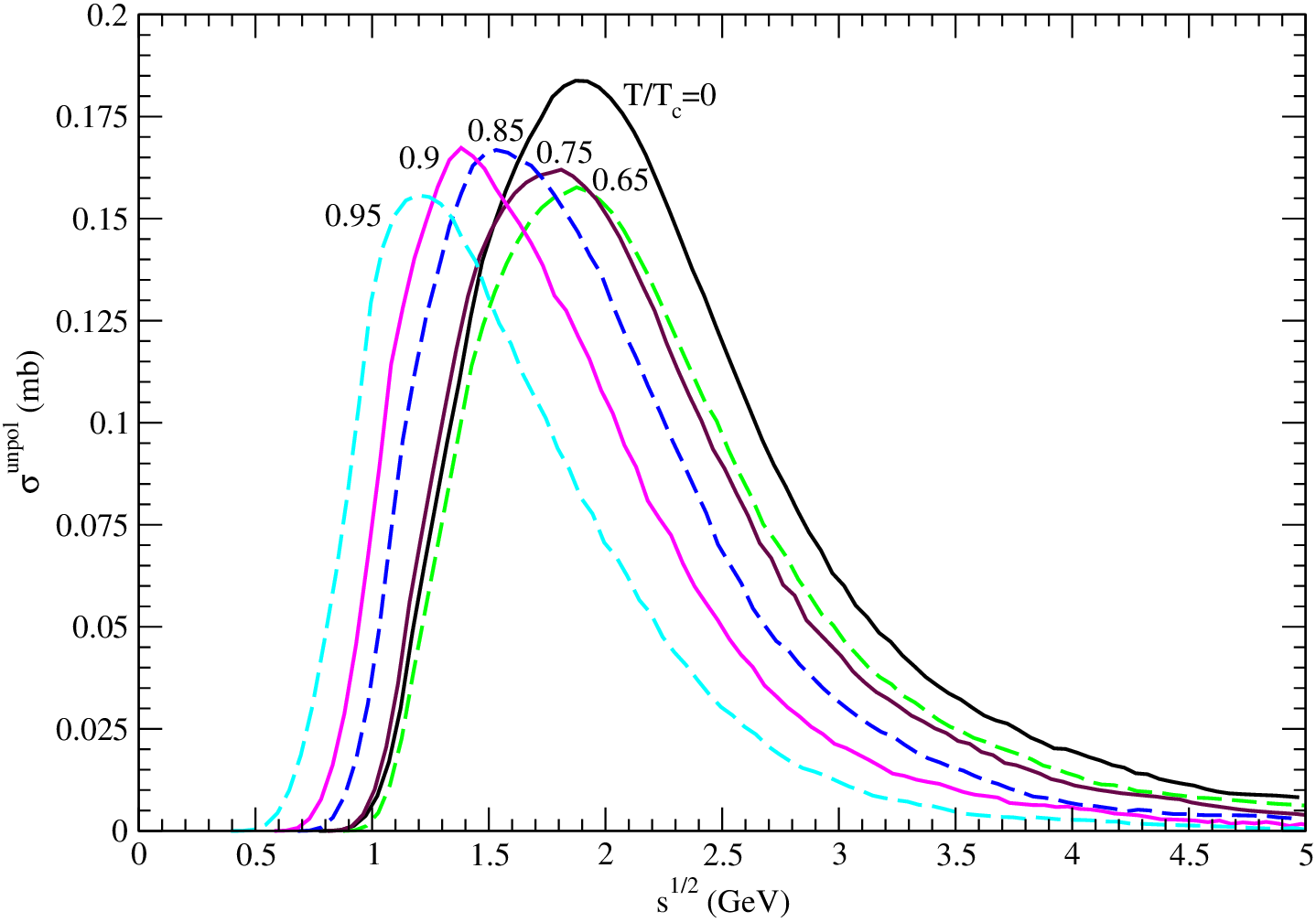}
\caption{Cross sections for $\pi K \to \pi \pi K$ for $I=3/2$ and
$I_{\pi K}^f=3/2$ at various temperatures.}
\label{fig6}
\end{figure}

\newpage
\begin{figure}[htbp]
  \centering
    \includegraphics[scale=0.6]{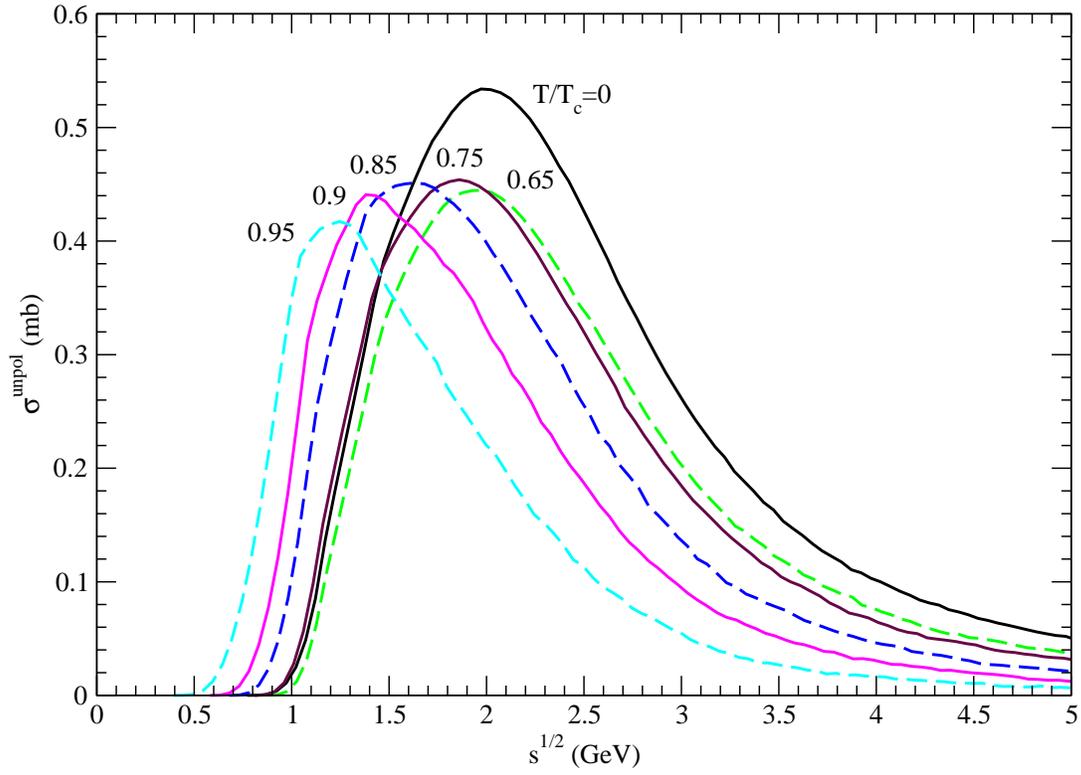}
\caption{Cross sections for $\pi K \to \pi \pi K$ for $I=3/2$ and
$I_{\pi K}^f=1/2$ at various temperatures.}
\label{fig7}
\end{figure}

\newpage
\begin{figure}[htbp]
  \centering
    \includegraphics[scale=0.6]{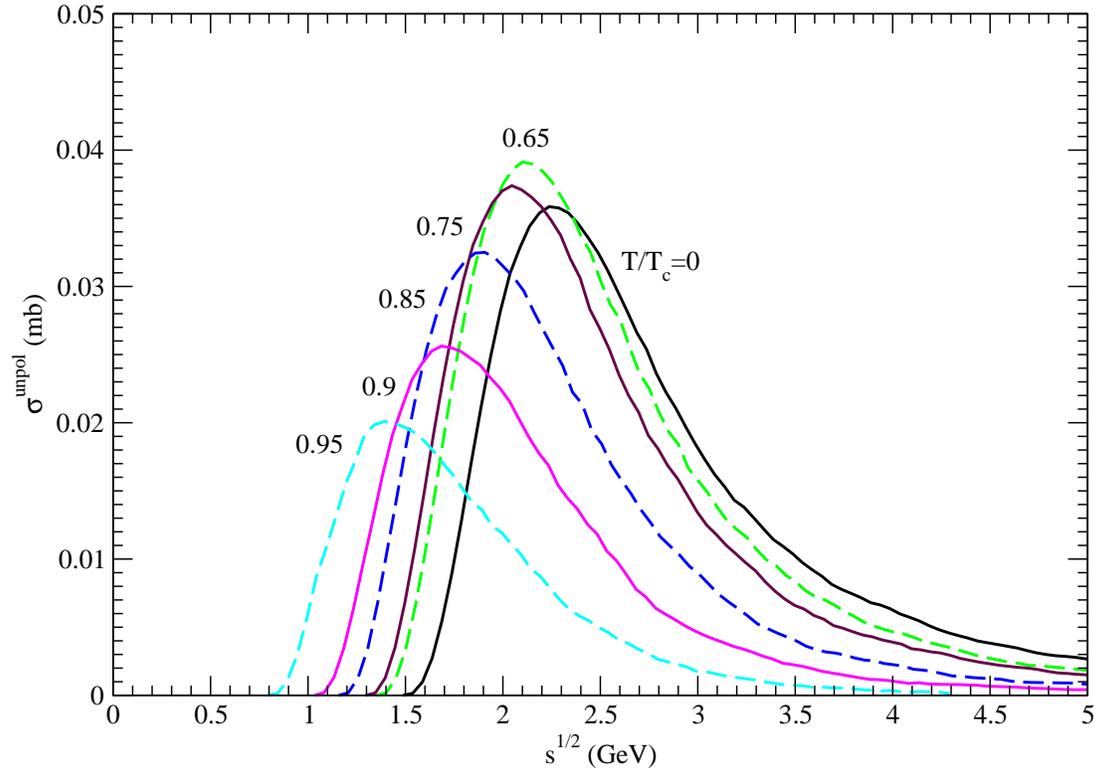}
\caption{Cross sections for $\pi K \to KK\bar K$ for $I=3/2$ and
$I_{KK}^f=1$ at various temperatures.}
\label{fig8}
\end{figure}

\newpage
\begin{figure}[htbp]
  \centering
    \includegraphics[scale=0.6]{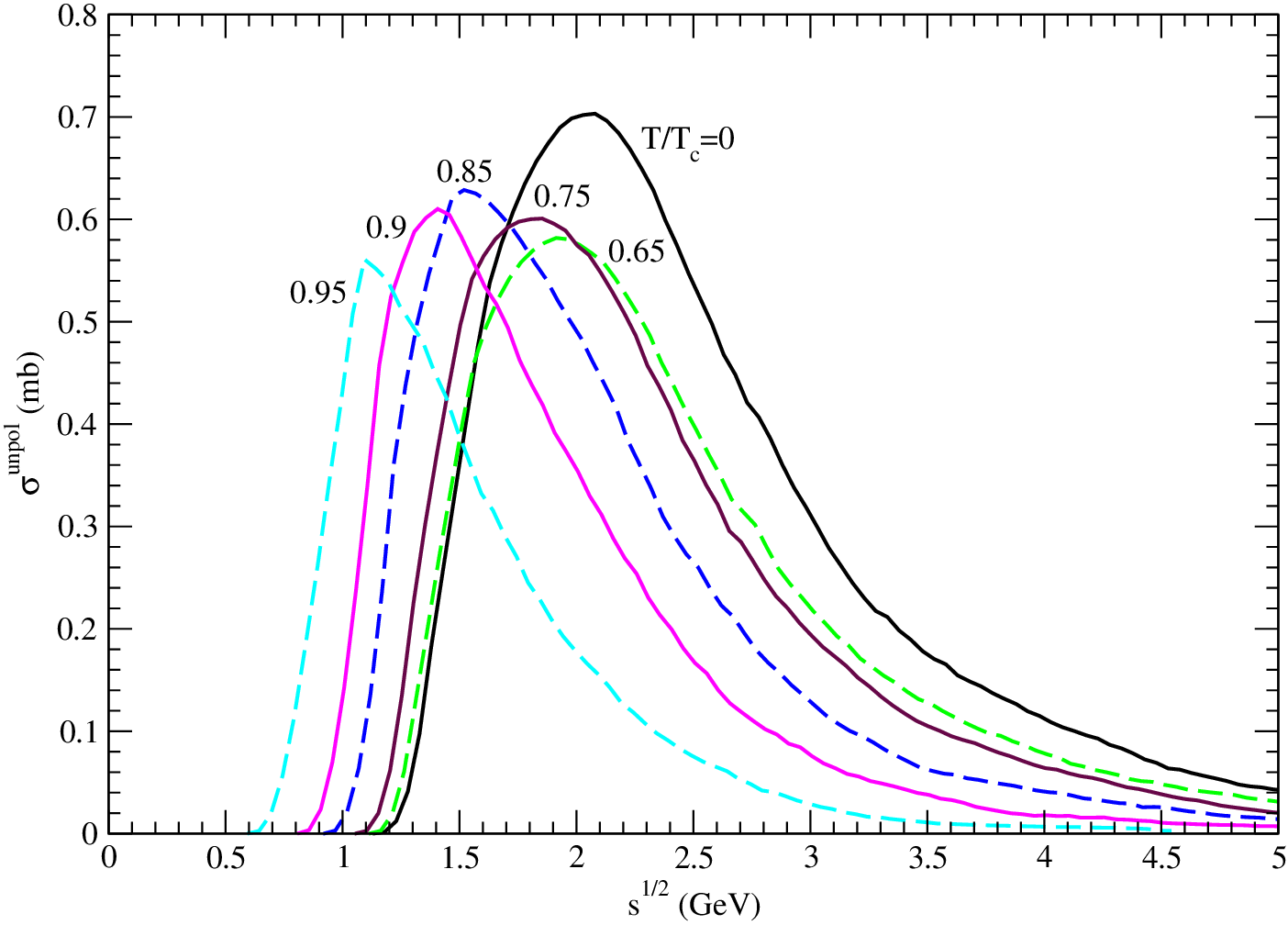}
\caption{Cross sections for $KK \to \pi KK$ for $I=1$ and
$I_{\pi K}^f=3/2$ at various temperatures.}
\label{fig9}
\end{figure}

\newpage
\begin{figure}[htbp]
  \centering
    \includegraphics[scale=0.6]{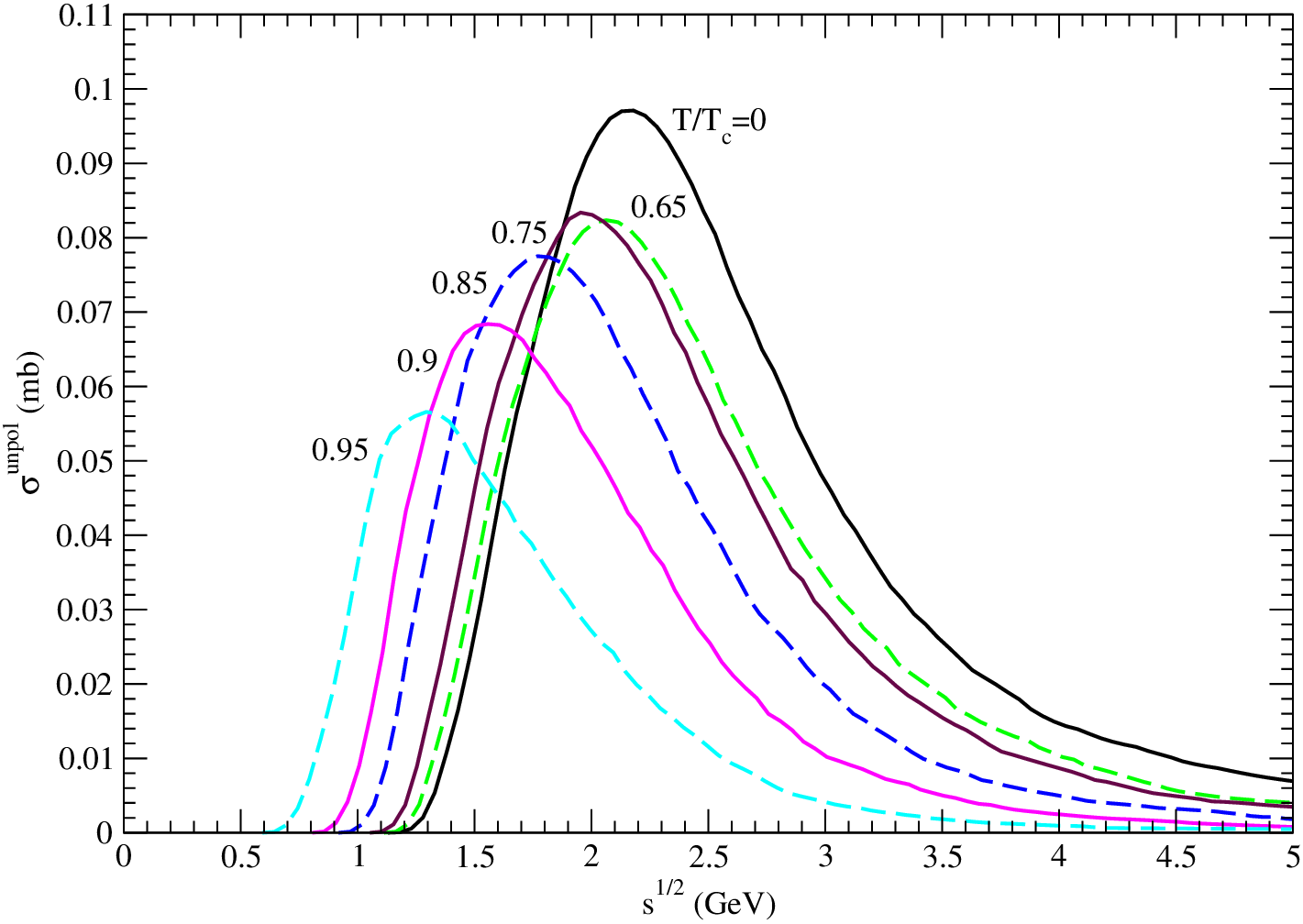}
\caption{Cross sections for $KK \to \pi KK$ for $I=1$ and
$I_{\pi K}^f=1/2$ at various temperatures.}
\label{fig10}
\end{figure}

\newpage
\begin{figure}[htbp]
  \centering
    \includegraphics[scale=0.6]{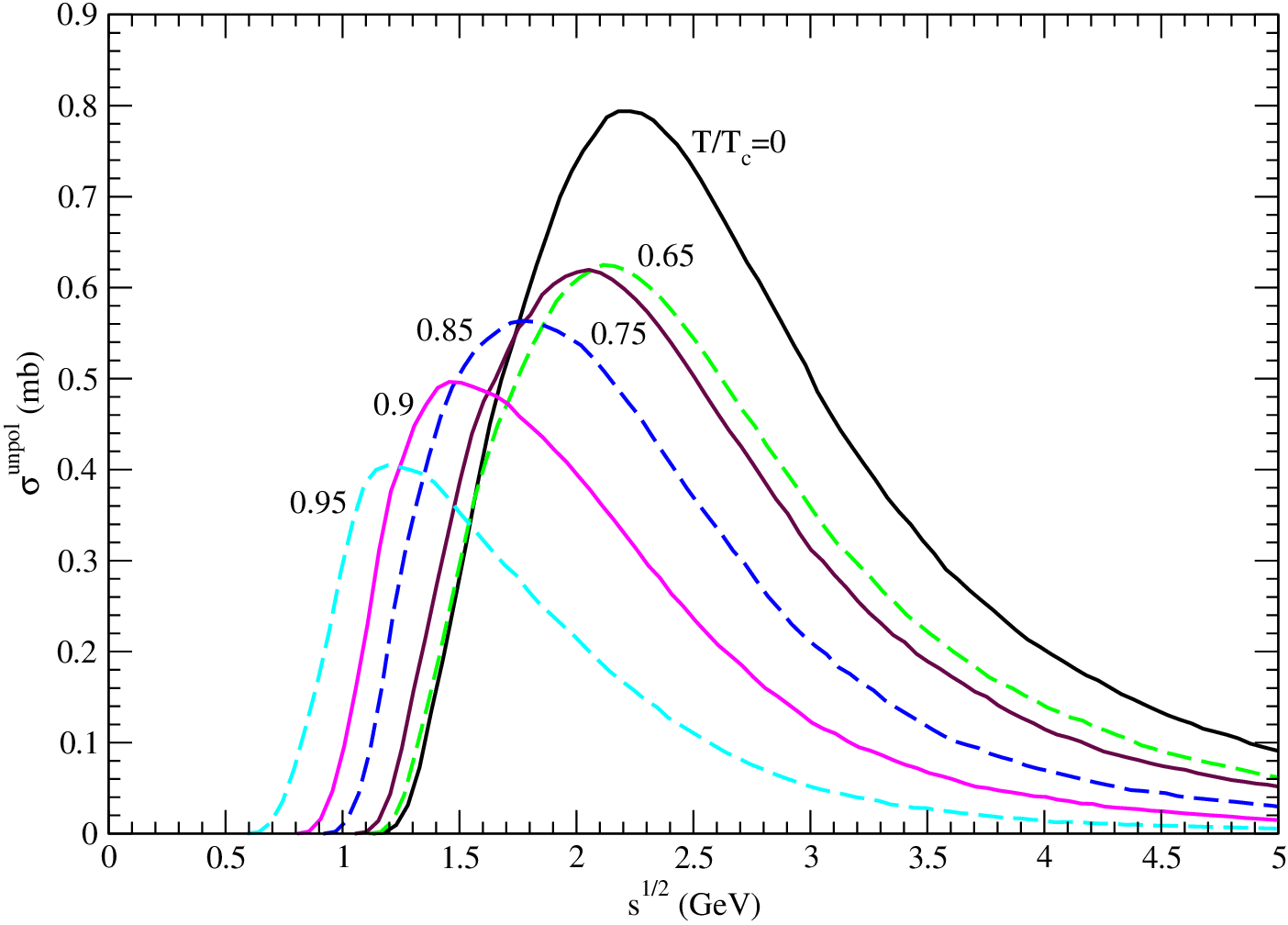}
\caption{Cross sections for $K\bar{K} \to \pi K\bar{K}$ for $I=1$ and
$I_{\pi \bar{K}}^f=3/2$ at various temperatures.}
\label{fig11}
\end{figure}

\clearpage
\begin{table}
\caption{\label{table1} Flavor matrix elements in the second column are part 
of ${\cal M}_{\rm D_1f}$, ${\cal M}_{\rm D_2f}$, ${\cal M}_{\rm D_3f}$, and
${\cal M}_{\rm D_4f}$, and in the third column of 
${\cal M}_{\rm D_5f}$, ${\cal M}_{\rm D_6f}$, ${\cal M}_{\rm D_7f}$, and
${\cal M}_{\rm D_8f}$.}
\begin{tabular}{ccc}
\hline
$I=2~I_{\pi K}^{\rm f}=\frac{3}{2}~\pi\pi \to \pi K\bar{K}$ 
& 1 & 1 \\
$I=1~I_{\pi K}^{\rm f}=\frac{3}{2}~\pi\pi \to \pi K\bar{K}$ 
& $\frac {\sqrt 3}{3}$ & $-\frac {\sqrt 3}{3}$ \\
$I=1~I_{\pi K}^{\rm f}=\frac{1}{2}~\pi\pi \to \pi K\bar{K}$ 
& $\frac {1}{\sqrt 6}$ & $\frac {1}{\sqrt 6}$ \\
$I=0~I_{\pi K}^{\rm f}=\frac{1}{2}~\pi\pi \to \pi K\bar{K}$ 
& $\frac {1}{2}$ & $-\frac {1}{2}$ \\
$I=\frac{3}{2}~I_{\pi K}^{\rm f}=\frac{3}{2}~\pi K \to \pi \pi K$ 
& 0 & $\sqrt{\frac {5}{6}}$ \\
$I=\frac{1}{2}~I_{\pi K}^{\rm f}=\frac{3}{2}~\pi K \to \pi \pi K$ 
& 0 & $\frac {2}{\sqrt 3}$ \\
$I=\frac{3}{2}~I_{\pi K}^{\rm f}=\frac{1}{2}~\pi K \to \pi \pi K$ 
& $-\sqrt{\frac {3}{2}}$ & $-\frac {1}{\sqrt 6}$ \\
$I=\frac{1}{2}~I_{\pi K}^{\rm f}=\frac{1}{2}~\pi K \to \pi \pi K$ 
& $-\frac {1}{2}\sqrt {\frac {3}{2}}$ & $-\frac {1}{2\sqrt 6}$ \\
$I=\frac{3}{2}~I_{KK}^{\rm f}=1~\pi K \to KK\bar{K}$ 
& 0 & 1 \\
$I=\frac{1}{2}~I_{KK}^{\rm f}=1~\pi K \to KK\bar{K}$ 
& 0 & $-\frac {1}{2}$ \\
$I=\frac{1}{2}~I_{KK}^{\rm f}=0~\pi K \to KK\bar{K}$ 
& 0 & $\frac {\sqrt 3}{2}$ \\
$I=1~I_{\pi K}^{\rm f}=\frac{3}{2}~KK \to \pi KK$ 
& $-\frac {2}{\sqrt 3}$ & $-\frac {2}{\sqrt 3}$ \\
$I=1~I_{\pi K}^{\rm f}=\frac{1}{2}~KK \to \pi KK$ 
& $-\frac {1}{\sqrt 6}$ & $\frac {1}{\sqrt 6}$ \\
$I=0~I_{\pi K}^{\rm f}=\frac{1}{2}~KK \to \pi KK$ 
& $-\frac {3}{\sqrt 6}$ & $-\frac {3}{\sqrt 6}$ \\
$I=1~I_{\pi \bar{K}}^{\rm f}=\frac{3}{2}~K\bar{K} \to \pi K\bar{K}$ 
& 0 & $-\frac {2}{\sqrt 3}$ \\
$I=1~I_{\pi \bar{K}}^{\rm f}=\frac{1}{2}~K\bar{K} \to \pi K\bar{K}$ 
& 0 & $\frac {\sqrt 6}{3}$ \\
$I=0~I_{\pi \bar{K}}^{\rm f}=\frac{1}{2}~K\bar{K} \to \pi K\bar{K}$ 
& 0 & 0 \\
\hline
\end{tabular}
\end{table}

\newpage
\begin{table}
\caption{\label{table2} Spin matrix elements in
${\cal M}_{\rm D_1}$, ${\cal M}_{\rm D_2}$, ${\cal M}_{\rm D_3}$, 
${\cal M}_{\rm D_4}$, ${\cal M}_{\rm D_5}$, ${\cal M}_{\rm D_6}$,
${\cal M}_{\rm D_7}$, and ${\cal M}_{\rm D_8}$,
which are shown from the second to ninth columns, respectively.
The initial spin state is
$\phi_{\rm iss}=\chi_{S_A S_{Az}} \chi_{S_B S_{Bz}}$, and
the final spin state $\phi_{\rm fss}=\chi_{S_{C_1}S_{C_1z}} 
\chi_{S_{C_2}S_{C_2z}} \chi_{S_{C_3}S_{C_3z}}$. The $z$ components of the meson
spins are $S_{Az}=S_{Bz}=S_{C_1z}=S_{C_2z}=S_{C_3z}=0$.}
\begin{tabular}{ccccccccc}
\hline
$\phi_{\rm fss}^+ \phi_{\rm iss}$ 
& 0 & 0 & 0 & 0 & 0 & 0 & 0 & 0 \\
$\phi_{\rm fss}^+ \sigma_1(34) \phi_{\rm iss}$ 
& 0 & 0 & 0 & 0 & 0 & 0 & 0 & 0 \\
$\phi_{\rm fss}^+ \sigma_2(34) \phi_{\rm iss}$ 
& $-\frac {1}{2\sqrt 2}i$ & $-\frac {1}{2\sqrt 2}i$ & $-\frac {1}{2\sqrt 2}i$ 
& $-\frac {1}{2\sqrt 2}i$ & $-\frac {1}{2\sqrt 2}i$ & $-\frac {1}{2\sqrt 2}i$
& $-\frac {1}{2\sqrt 2}i$ & $-\frac {1}{2\sqrt 2}i$ \\
$\phi_{\rm fss}^+ \sigma_3(34) \phi_{\rm iss}$ 
& 0 & 0 & 0 & 0 & 0 & 0 & 0 & 0 \\
$\phi_{\rm fss}^+ \sigma_1 \phi_{\rm iss}$ 
& 0 & 0 & 0 & 0 & 0 & 0 & 0 & 0 \\
$\phi_{\rm fss}^+ \sigma_2 \phi_{\rm iss}$
& $-\frac {1}{2\sqrt 2}i$ & $-\frac {1}{2\sqrt 2}i$ & $-\frac {1}{2\sqrt 2}i$ 
& $-\frac {1}{2\sqrt 2}i$ & $-\frac {1}{2\sqrt 2}i$ & $-\frac {1}{2\sqrt 2}i$
& $-\frac {1}{2\sqrt 2}i$ & $-\frac {1}{2\sqrt 2}i$ \\
$\phi_{\rm fss}^+ \sigma_3 \phi_{\rm iss}$ 
& 0 & 0 & 0 & 0 & 0 & 0 & 0 & 0 \\
\hline
$\phi_{\rm fss}^+ \sigma_1(34) \sigma_1 \phi_{\rm iss}$
& 0 & 0 & 0 & 0 & 0 & 0 & 0 & 0 \\
$\phi_{\rm fss}^+ \sigma_1(34) \sigma_2 \phi_{\rm iss}$
& 0 & 0 & 0 & 0 & 0 & 0 & 0 & 0 \\
$\phi_{\rm fss}^+ \sigma_1(34) \sigma_3 \phi_{\rm iss}$
& $\frac {1}{2\sqrt 2}$ & $-\frac {1}{2\sqrt 2}$ & $\frac {1}{2\sqrt 2}$ 
& $-\frac {1}{2\sqrt 2}$ & $\frac {1}{2\sqrt 2}$ & $-\frac {1}{2\sqrt 2}$
& $\frac {1}{2\sqrt 2}$ & $-\frac {1}{2\sqrt 2}$ \\
$\phi_{\rm fss}^+ \sigma_2(34) \sigma_1 \phi_{\rm iss}$
& 0 & 0 & 0 & 0 & 0 & 0 & 0 & 0 \\
$\phi_{\rm fss}^+ \sigma_2(34) \sigma_2 \phi_{\rm iss}$
& 0 & 0 & 0 & 0 & 0 & 0 & 0 & 0 \\
$\phi_{\rm fss}^+ \sigma_2(34) \sigma_3 \phi_{\rm iss}$
& 0 & 0 & 0 & 0 & 0 & 0 & 0 & 0 \\
$\phi_{\rm fss}^+ \sigma_3(34) \sigma_1 \phi_{\rm iss}$
& $-\frac {1}{2\sqrt 2}$ & $\frac {1}{2\sqrt 2}$ & $-\frac {1}{2\sqrt 2}$ 
& $\frac {1}{2\sqrt 2}$ & $-\frac {1}{2\sqrt 2}$ & $\frac {1}{2\sqrt 2}$
& $-\frac {1}{2\sqrt 2}$ & $\frac {1}{2\sqrt 2}$ \\
$\phi_{\rm fss}^+ \sigma_3(34) \sigma_2 \phi_{\rm iss}$
& 0 & 0 & 0 & 0 & 0 & 0 & 0 & 0 \\
$\phi_{\rm fss}^+ \sigma_3(34) \sigma_3 \phi_{\rm iss}$
& 0 & 0 & 0 & 0 & 0 & 0 & 0 & 0 \\
\hline
\end{tabular}
\end{table}

\newpage
\begin{table*}[htbp]
\caption{\label{table3}Values of the parameters. $a_1$ and $a_2$ are
in units of millibarns; $b_1$, $b_2$, $d_0$, and $\sqrt{s_{\rm z}}$ are
in units of GeV; $e_1$ and $e_2$ are dimensionless.}
\tabcolsep=5pt
\begin{tabular}{cccccccccc}
  \hline
  \hline
Reactions & $T/T_{\rm c} $ & $a_1$ & $b_1$ & $e_1$ & $a_2$ & $b_2$ & $e_2$ &
$d_0$ & $\sqrt{s_{\rm z}} $\\
\hline
 $I=2~I_{\pi K}^{\rm f}=\frac{3}{2}~\pi\pi \to \pi K\bar{K}$
  &  0     & 0.1  & 0.81  & 3.81 & 0.05 & 1.4   & 5.06  & 0.95 & 5.83\\
  &  0.65  & 0.09 & 0.72  & 3.24 & 0.06 & 1.19  & 4.12  & 0.9  & 5.82\\
  &  0.75  & 0.09 & 0.71  & 3.37 & 0.06 & 1.18  & 4.16  & 0.9  & 5.6\\
  &  0.85  & 0.08 & 0.64  & 3.69 & 0.07 & 1.13  & 4.59  & 0.85 & 5.06\\
  &  0.9   & 0.07 & 0.62  & 3.57 & 0.06 & 1.07  & 4.31  & 0.8  & 5.01\\
  &  0.95  & 0.07 & 0.6   & 4.95 & 0.06 & 1.05  & 4.68  & 0.7  & 4.26\\
  \hline
 $I=1~I_{\pi K}^{\rm f}=\frac{3}{2}~\pi\pi \to \pi K\bar{K}$
  &  0     & 0.03  & 1.04  & 4.88 & 0.02  & 1.27  & 2.9  & 1.1  & 6.08\\
  &  0.65  & 0.03  & 1.01  & 3.55 & 0.02  & 1.09  & 3.14 & 1.05 & 5.52\\
  &  0.75  & 0.03  & 0.99  & 4.5  & 0.02  & 1.11  & 2.64 & 1.05 & 5.37\\
  &  0.85  & 0.03  & 0.84  & 3.8  & 0.02  & 1.33  & 5.97 & 1.05 & 5.15\\
  &  0.9   & 0.02  & 0.81  & 3.5  & 0.02  & 1.19  & 5.06 & 1    & 5.08\\
  &  0.95  & 0.02  & 0.91  & 3.79 & 0.01  & 1.09  & 3.87 & 1    & 4.63\\
  \hline
 $I=\frac{3}{2}~I_{\pi K}^{\rm f}=\frac{3}{2}~\pi K \to \pi \pi K$
  &  0     & 0.13  & 0.81  & 5.5  & 0.11  & 1.52  & 6.67 & 1.1  & 6.36\\
  &  0.65  & 0.1   & 0.67  & 4.29 & 0.1   & 1.31  & 5.2  & 1    & 5.87\\
  &  0.75  & 0.1   & 0.69  & 4.51 & 0.1   & 1.24  & 4.3  & 1    & 5.35\\
  &  0.85  & 0.12  & 0.68  & 5.27 & 0.1   & 1.28  & 5.72 & 0.85 & 5.15\\
  &  0.9   & 0.11  & 0.68  & 5.36 & 0.1   & 1.19  & 5.56 & 0.8  & 4.89\\
  &  0.95  & 0.12  & 0.78  & 5.66 & 0.06  & 1.26  & 5    & 0.8  & 4.38\\
  \hline
 $I=\frac{3}{2}~I_{\pi K}^{\rm f}=\frac{1}{2}~\pi K \to \pi \pi K$
  &  0     & 0.35 & 0.84  & 5.17 & 0.34 & 1.66  & 4.84 & 1.2    & 5.8\\
  &  0.65  & 0.29 & 0.73  & 3.94 & 0.27 & 1.5   & 4.18 & 1.1    & 6.62\\
  &  0.75  & 0.3  & 0.7   & 4.62 & 0.3  & 1.44  & 4.27 & 1.05   & 5.97\\
  &  0.85  & 0.3  & 0.69  & 5.04 & 0.3  & 1.36  & 4.49 & 0.95   & 5.82\\
  &  0.9   & 0.29 & 0.67  & 5.51 & 0.29 & 1.3   & 4.81 & 0.85   & 5.49\\
  &  0.95  & 0.27 & 0.77  & 6.41 & 0.21 & 1.23  & 3.59 & 0.85   & 5.24\\
  \hline
  \hline
\end{tabular}
\end{table*}

\newpage
\begin{table*}[htbp]
\caption{\label{table4}The same as Table 3, but for four other reactions}
\tabcolsep=5pt
\begin{tabular}{cccccccccc}
  \hline
  \hline
Reactions & $T/T_{\rm c} $ & $a_1$ & $b_1$ & $e_1$ & $a_2$ & $b_2$ & $e_2$ &
$d_0$ & $\sqrt{s_{\rm z}} $\\
\hline
 $I=\frac{3}{2}~I_{KK}^{\rm f}=1~\pi K \to KK\bar{K}$
  &  0     & 0.02 & 0.6   & 3.92 & 0.02 & 1.08  & 2.82  & 0.75 & 6\\
  &  0.65  & 0.02 & 0.68  & 3.75 & 0.02 & 0.95  & 2.15  & 0.75 & 5.85\\
  &  0.75  & 0.03 & 0.69  & 3.01 & 0.01 & 1.19  & 2.69  & 0.75 & 5.64\\
  &  0.85  & 0.02 & 0.52  & 3.65 & 0.02 & 1.03  & 4.07  & 0.75 & 5.6\\
  &  0.9   & 0.01 & 0.43  & 3.19 & 0.02 & 0.87  & 3.57  & 0.65 & 5.49\\
  &  0.95  & 0.01 & 0.6   & 3.22 & 0.01 & 0.75  & 1.87  & 0.6  & 4.26\\
  \hline
 $I=1~I_{\pi K}^{\rm f}=\frac{3}{2}~KK \to \pi KK$
  &  0     & 0.46  & 0.65  & 3.43 & 0.38  & 1.26  & 3.11 & 0.95 & 6.15\\
  &  0.65  & 0.38  & 0.64  & 2.63 & 0.27  & 1.16  & 2.52 & 0.8  & 5.91\\
  &  0.75  & 0.4   & 0.55  & 3.08 & 0.35  & 1.14  & 3.47 & 0.8  & 5.54\\
  &  0.85  & 0.41  & 0.47  & 4.2  & 0.41  & 1.03  & 3.81 & 0.6  & 5.52\\
  &  0.9   & 0.4   & 0.45  & 4.69 & 0.39  & 0.94  & 3.93 & 0.6  & 5.49\\
  &  0.95  & 0.38  & 0.51  & 4.4  & 0.25  & 0.9   & 3.67 & 0.5  & 4.3\\
  \hline
 $I=1~I_{\pi K}^{\rm f}=\frac{1}{2}~KK \to \pi KK$
  &  0     & 0.07  & 0.92  & 3.64 & 0.03  & 1.44  & 3.16 & 1.05 & 6.49\\
  &  0.65  & 0.07  & 0.83  & 3.13 & 0.02  & 1.56  & 4.55 & 0.95 & 6.72\\
  &  0.75  & 0.05  & 0.73  & 3.22 & 0.04  & 1.21  & 3.54 & 0.9  & 6.27\\
  &  0.85  & 0.05  & 0.67  & 3.63 & 0.04  & 1.15  & 3.72 & 0.85 & 5.52\\
  &  0.9   & 0.04  & 0.54  & 4.52 & 0.05  & 1.04  & 4.58 & 0.75 & 5.08\\
  &  0.95  & 0.04  & 0.58  & 4.41 & 0.03  & 1.08  & 5.14 & 0.7  & 4.53\\
  \hline
 $I=1~I_{\pi \bar{K}}^{\rm f}=\frac{3}{2}~K\bar{K} \to \pi K\bar{K}$
  &  0     & 0.52 & 0.83  & 3.01 & 0.38 & 1.61  & 3.38 & 1.05   & 6.3\\
  &  0.65  & 0.44 & 0.84  & 2.52 & 0.23 & 1.51  & 2.54 & 1      & 6.18\\
  &  0.75  & 0.42 & 0.73  & 2.72 & 0.3  & 1.45  & 3.27 & 1      & 6.05\\
  &  0.85  & 0.37 & 0.57  & 3.56 & 0.38 & 1.26  & 3.7  & 0.85   & 5.66\\
  &  0.9   & 0.34 & 0.53  & 3.81 & 0.32 & 1.16  & 3.7  & 0.65   & 5.64\\
  &  0.95  & 0.28 & 0.56  & 4.01 & 0.22 & 1.09  & 3.68 & 0.6    & 5.22\\
  \hline
  \hline
\end{tabular}
\end{table*}

\end{document}